\documentclass[useAMS,usenatbib,usegraphicx]{mn2e}
\usepackage{times}
\usepackage{amssymb}
\usepackage{graphicx}
\usepackage{longtable}

\newcommand{\kms}{$\rm km\,s^{-1}$}
\newcommand{\am}{$^{\prime}$}

\newcommand{\gsh}{GSH\,117.8+1.5-35}

\newcommand{\hi}{\mbox{H\,\sc{i}}}
\newcommand{\hii}{\mbox{H\,\sc{ii}}}

\title[Unveiling the birth and evolution of Sh2-173]{Unveiling the birth and evolution of the \hii\, region Sh2-173}

\author[S. Cichowolski et al.] {S.~Cichowolski,$^1$\thanks{Member of
    the Carrera del Investigador Cient\'{\i}fico of CONICET,
    Argentina}  G.A.~Romero,$^{2, 3}$
    M.E.~Ortega,$^1$\thanks{Doctoral Fellow of CONICET, Argentina} 
C.E.~Cappa, $^2$\footnotemark[1]
\newauthor 
and  
 J.~Vasquez,$^2$
 \thanks{Post-Doc Fellow of CONICET, Argentina}\\
$^1$ Instituto de Astronom\'{\i}a y F\'{\i}sica del Espacio (IAFE), CC 67, Suc. 28, 1428 Buenos Aires, Argentina\\
 $^{2}$ Instituto Argentino de Radioastronom\'{\i}a  (CCT-La Plata, CONICET), CC 5, 1894 Villa Elisa,
  Argentina and Facultad de Ciencias \\ Astron\'omicas y
  Geof\'{\i}sicas, Universidad Nacional de La Plata, Paseo del
  Bosque s/n, 1900 La Plata, Argentina\\
$^{3}$ Departamento de F\'{\i}sica y Astronom\'{\i}a, Facultad de Ciencias,
Universidad de Valparaiso, Valparaiso, Chile
}

\date{}

\begin{document}
\maketitle

\begin{abstract}

Based on a multiwavelength study, the interstellar medium around the \hii\ region Sh2-173 has been analyzed. The ionized region is clearly detected in the optical and in the radio continuum images. 
The analysis of the \hi\ data shows a region of low emissivity that has an excellent morphological correlation with the radio continuum emission. 
The \hii\ region is partially  bordered by a photodissociation region, which, in turn, is encircled  by a molecular structure. The \hi\ and CO structures  related to Sh2-173 are observed in the velocity ranges from --25 to --31\kms, and from --27 to --39 \kms, respectively.
Taking into account the presence of noncircular motions in the Perseus spiral arm, together with previous distance estimates for the region, 
 we adopt a distance of 2.5 $\pm$ 0.5 kpc for Sh2-173. 
Seven hot stars were identified in the field of Sh2-173, being only one an O-type star. The amount of energetic photons emitted by this star is enough to keep the region ionized and heat the dust.
Given that an expanding \hii\ region may trigger star formation, a search for YSO candidates was made using different infrared point source catalogues. A population of 46 YSO candidates was identified projected onto the molecular clouds.\\
On the other hand, Sh2-173 is located in a dense edge of a large ($\sim5^{\circ}$) \hi\ shell, \gsh.  
The possibility for Sh2-173 of being part of a hierarchical system of three generations is suggested. In this scenario, the large \hi\ shell, which was probably originated due to  the action of Cas\,OB5, would have triggered the formation of Sh2-173, which, in turn, is triggering new stars in its surrounding molecular cloud. 
To test this hypothesis, the ages of both, the \hii\ region and the large shell, were estimated and compared. We concluded that Sh2-173 is a young \hii\ region of about  0.6 - 1.0  Myr old. As for the large shell we obtained a dynamical age of 5 $\pm$ 1 Myr.
 These age estimates, together with the relative location of the different structures, support the hypothesis that Sh2-173 is part of a hierarchical system.

\end{abstract} 

\begin{keywords}
\ ISM: individual:Sh2-173\ -- \ ISM: kinematics and dynamics\ -- \ \hii\, regions\ -- \ stars: formation. 

\end{keywords}

\section{Introduction}\label{intro}

Massive stars affect their environment by ionizing radiation, stellar winds, and, finally, supernova explosions.
These stars produce intense extreme ultraviolet ($h\nu > 13.6$ eV) radiation, creating  \hii\ regions around them, and intense far 
ultraviolet ($4.5 < h\nu < 13.6$ eV) radiation that can penetrate into regions of molecular gas, dissociating it. They also have  powerful
 winds which sweep up the surrounding gas creating what is known  as  an interstellar bubble, i.e. a minimum in the gas distribution 
characterized by a low volume density and a high temperature, surrounded by an expanding envelope \citep{wea77}.
In the case of OB associations, their collective effect has a huge impact in the surrounding interstellar medium (ISM), creating large 
structures of about hundred parsecs. Winds and supernova explosions transfer to the ISM vast amounts of mechanical energy, generating 
the compression of nearby molecular clouds. 
The effects of the massive stars are mostly destructive in their immediate neighborhood, since they tend to disrupt the parental molecular 
cloud. 
However, a little further away, massive stars stimulate star formation.
 A number of papers have been devoted to investigate star formation in the
dense molecular clouds in the  surroundings of \hii\ regions, both from a
theoretical and an observational point of view.

Different processes  can trigger stellar formation at the periphery of
\hii\ regions \citep{elm77, kle80, san82, lef94, whi94, lef95, deh05, dal07}.
Star formation can be induced in the environs of massive stars
due to the action of expanding shocks which modify the surrounding material,
either favouring the formation of cometary globules in preexistent clumps ("Radiative
Driven Implosion" model, RDI) or sweeping up the surrounding material into a
collected layer, which latter becomes gravitationally unstable generating massive fragments
("collect and collapse" model). Both processes, which may act simultaneously in
an \hii\ region, lead to the formation of a new generation of stars.
Observational evidence of both (the RDI and "collect and collapse")
processes  has been found in a number of \hii\ regions \citep[e.g.][]{tho04, deh08, cap08, kir08}.

The goal of this paper is to achieve a better understanding of the phenomena associated with the interaction between young massive stars and their surrounding interstellar medium and, specifically, to attempt to shed some light on the process of triggered star formation.
To carry out this study a multiwavelength approach is essential in order to obtain a complete
picture of the different components of the ISM, the young stellar
population, and the interactions among them.

\begin{figure}
\includegraphics[]{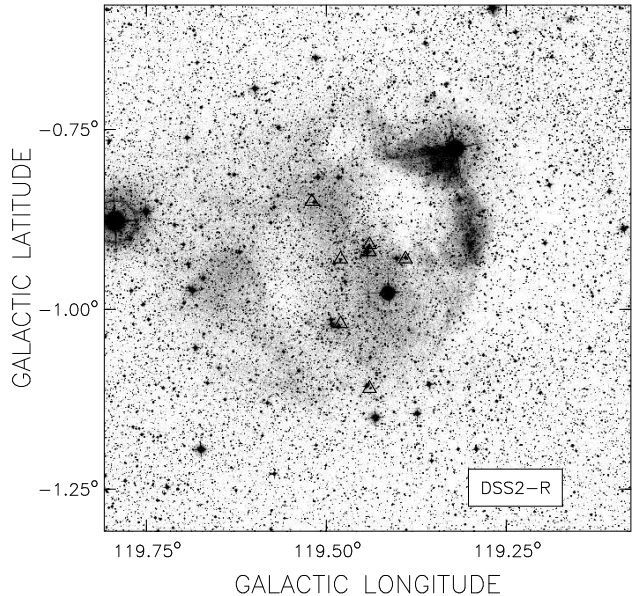}
\caption{DSS2-R optical image of the \hii\ region Sh2-173. The small
  triangles indicate the position of the seven OB stars
  listed in Table \ref{tabest}. }
\label{optico}
\end{figure}

In this work we investigate the environs of the \hii\ region Sh2-173
by analyzing the distribution of the ionized and neutral material, and that of the interstellar dust.
Sh2-173 is an optically visible \hii\ region \citep{sha59} located at
({\it l, b}) = (119\fdg4, --0\fdg94) in the Perseus spiral arm. 
The DSS2-R image of Sh2-173 is displayed in Fig. \ref{optico}. 
The region is approximately 
circular in shape with a diameter of about 30\am.
An arc-like structure of enhanced optical emission is evident on the north-west side of the region,
while fainter emission  is present at higher  galactic longitudes and lower galactic latitudes. 
Inside the ionized region, two voids lacking 
optical emission are evident  at ({\it l, b}) = 
(119\fdg4, --0\fdg85) and  ({\it l, b}) =
(119\fdg55, --0\fdg95).
\citet*{rus07} identified seven hot stars in the field of Sh2-173,
which are listed in Table\ref{tabest}, and indicated with small triangles in Fig. \ref{optico}. All of the stars but one  
are projected close to the center of the region. ALS\,6155, on the contrary,
is seen projected on its southeast border. 

The distance to Sh2-173 has been estimated by several authors using independent methods.
A spectrophotometric distance of $ 2.7 \pm 0.9$ kpc was derived  by
\citet{geo75}, while \citet{rus07} estimated a distance of $
3.12 \pm 0.34$ kpc based on UBV photometric and spectroscopic data of the
exciting stars  (see Table \ref{tabest}).
Using  a new method for determination of distances, which is  based on \hi\ column densities and is independent of Galactic kinematics, 
\citet{fos03} estimated for this \hii\ region a distance of $ 2.2 \pm 0.4$ kpc.
\citet*{bli82} in their CO survey towards \hii\ regions found  molecular gas related to Sh2-173 at --34.5 \kms\ (all the velocities are 
referred to the LSR).
This velocity coincides with  the radial velocity of the ionized gas derived
from Fabry-Perot H$\alpha$ observations, $-34.3 \pm 0.3$
\kms\, \citep*{fic90}.
It is well known that  noncircular motions on a large scale are present in the Perseus spiral arm \citep{bra93}.
An inspection of  Fig. 2b of \citet{bra93}, which takes into account noncircular 
motions, suggests  for a velocity of about --34 \kms\,, a kinematical distance of 1.8-2 kpc.
Thus, the distance of Sh2-173 is between 1.8 and 3.1 kpc. As a working  hypothesis, we will adopt a distance of 2.5 $\pm$ 0.5 kpc for Sh2-173.

Sh2-173 is seen located  in the dense border of a large \hi\ shell 
 ($\sim 5^{\circ}$ diameter) that has been found in the Perseus spiral arm by \citet{fic86} using \citet{wea73} \hi\ data. 
\citet{fic86} suggested that the presence of SNR remnants and several \hii\ regions in the field indicates a great deal of star-formation 
activity going on nearby.
 \citet*{kis04} found a far infrared ring close to the \hi\ shell defined by \citet{fic86}. \citet{hey98} found a large void of CO emission 
centered at ({\it l, b}) = (117\fdg6, 0\fdg0), coincident with the \hi\ shell. 
Based on a multiwavelength study, \citet{moo03} determined the main parameters of the shell and associated it with the OB association 
Cas\,OB5 ($\equiv$ Rupr\,33).
According to \citet {gar92}, Cas\,OB5 has a zero-age main sequence (ZAMS) fairly good defined with a distance modulus DM = 11.5 mag 
(D $\sim$ 2 kpc). The diameter of the association is about 4$^{\circ}$, or  140 pc at 2 kpc. Its earliest star is HD\,108 (O6f). 
Among their members there are six O-type stars, being the others B stars \citep{gar92}.
A WN4 star (WR\,159) was found in the area of Cas\,OB5,  located at  ({\it l, b}) = (115\fdg78, +1\fdg24) \citep{neg03}. 

\begin{table}
\centering
\begin{minipage}{100mm}
\caption{Stellar parameters of the OB stars associated with Sh2-173.\label{tabest}}
\begin{tabular}{l c c c c}
\hline
Star & Spectral  & Galactic  &  Distance$^1$& Comments \\
& Type$^1$& coordinates &  [kpc] &\\
\hline
ALS6145 & B0.5V & 119\fdg39, --0\fdg93 &  $ 3.17 \pm 1.27 $  &\\
ALS6150 & B2IV & 119\fdg44, --0\fdg91 &  $ 3.78 \pm 1.89$ &\\
ALS6151 & O9V & 119\fdg44, --0\fdg92 &  $2.57 \pm 0.5$ & BD+60\,39\\
ALS6155 & B0.5V & 119\fdg44, --1\fdg11 &  $3.36 \pm 1.34$ & \\  
ALS6156 & B2V & 119\fdg48, --0\fdg93 &  $4.41 \pm 1.77 $ &\\
ALS6157 & B2IV & 119\fdg48, --1\fdg02 &  $ 2.27 \pm 1.14 $ & BD+60\,42 \\
ALS6158 & B1V & 119\fdg52, --0\fdg85 &  $ 4.02 \pm 0.66 $ &\\
\hline

\end{tabular}

\medskip
1- From \citet{rus07}.

\end{minipage}
\end{table}

\section{Data sets}

Radio continuum data at 408 and 1420 MHz, as well as 21-cm spectral
line data, were obtained using the Dominion Radio Astrophysical
Observatory (DRAO) interferometer as part of the Canadian Galactic
Plane Survey (CGPS) \citep{tay03}.  The CGPS data base also comprises
other data sets that have been reprojected and regridded to match the
DRAO images. Among them are the IRAS high resolution (HIRES) data
\citep{fow94} and the Five College Radio Astronomical Observatory
(FCRAO) CO Survey of the Outer Galaxy \citep{hey98}.

Radio continuum surveys at 325 MHz (WENSS) \citep{ren97}, 
2695 MHz \citep{fur90}, and 4850 MHz \citep{con94} were
also used in this work.  

Mid-infrared data were taken from the {\it MSX} Galactic Plane Survey \citep{pri01} and near-infrared data were obtained from the 2MASS All-Sky Point Source Catalog \citep{skr06}. 

Table \ref{tabobs} summarizes  the most relevant
observational parameters.

\begin{table}
\centering
\begin{minipage}{80mm}
\caption{Observational parameters. \label{tabobs}}
\begin{tabular}{l c}
\hline
\bf {\hi\ data:} & \\
Synthesized beam ($\Theta_l \times \Theta_b $)& 1\farcm1 $\times$ 1\farcm0\\
Position angle$^{\it a}$ & $97^\circ$\\
Velocity coverage & $-164$ to 44.7 \kms \\
Channel separation & 0.82 \kms \\
Velocity resolution & 1.3 \kms \\
Observed rms noise (single channel) & 1.3 K \\
\hline
\bf {Radio continuum at 408 MHz:} &\\
Synthesized beam ($\Theta_l \times \Theta_b $)& 3\farcm2 $\times$ 2\farcm8 \\
Position angle$^{\it a}$  & $95^\circ$\\
Observed rms noise & 0.5 K\\
\hline
\bf {Radio continuum at 1420 MHz:} &\\
Synthesized beam ($\Theta_l \times \Theta_b $ ) & 0\farcm93 $\times$ 0\farcm8 \\
Position angle$^{\it a}$  & $97^\circ$\\
Observed rms noise & 0.05 K \\
\hline
\bf {CO data:} & \\
Angular resolution & 1\farcm67 \\
Velocity coverage & $-150.7$ to 38.9 \kms \\
Velocity resolution & 0.98 \kms \\
Channel separation  & 0.82 \kms \\
Observed rms noise (single channel)  & 0.2 K \\
\hline
\bf {Infrared data:} & \\
HIRES & \\
Angular resolution &  0\farcm5 - 2\am \\
MSX & \\
Angular resolution & 18\farcs4 \\
\hline
\end{tabular}
\medskip

{\it a}   Position angle measured CCW from the longitude axis.
\end{minipage}
\end{table}

\section{Sh2-173 and its local ISM}

\subsection{Ionized Gas}

\begin{figure}
\includegraphics[width=8 cm]{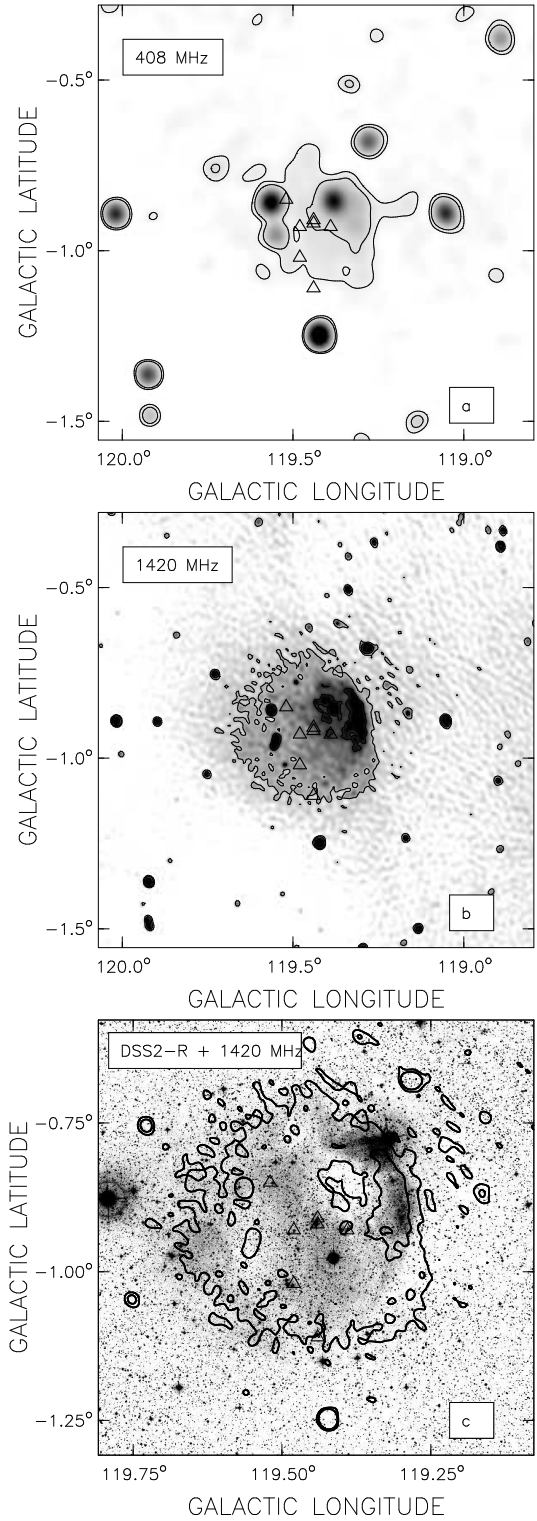}
\caption{Radio continuum images of the \hii\ region Sh2-173 obtained
  from  the CGPS. The small triangles indicate the position of
  the OB stars. ({\it a}) 408 MHz. Contours levels are at 58 and 62 K. 
({\it b}) 1420 MHz. Contour levels are
  at 5.7 and 6.8 K. ({\it c}) DSS2-R image (grey-scale) with the 1420 MHz radio continuum
  contours superimposed. The plotted area is smaller than the one shown in the other images. 
 }
\label{set-continuo}
\end{figure}

Figure \ref{set-continuo} shows the radio continuum emission
distribution at different wavelengths. 
 The upper panel shows the radio emission at 408 MHz, while 
the image at 1420 MHz is shown
in the middle panel.
Clearly, the
radio emission at 408 MHz is dominated by strong point
like sources. An arc of strong radio continuum emission overlaying a more extended
diffuse emission is observed at 1420 MHz. The radio emission extends
towards  higher galactic longitudes and lower galactic latitudes, slowly
decreasing in brightness. 
The bottom panel displays an
overlay of the emission at 1420 MHz ({\it line contours}) and the optical image ({\it grey-scale}).  
The arc-like structure is 
coincident with the brightest optical region. On the
contrary, the relatively strong radio emission near ({\it l, b}) =
(119\fdg4, --0\fdg85) is projected onto one of the regions lacking
optical emission.
The \hii\ region shows a similar morphology at 325, 2695, and 4850 MHz.

To elucidate the nature of the numerous point sources observed in the field, we have used the available 
catalogued flux densities (The Texas Survey [365 MHz, \citep{dou96}], WENSS [325 MHz, \citep{ren97}], NVSS [1420 MHz, \citep{con98}]). 
The derived  spectral indexes ($ S_\nu \sim \nu^\alpha$)
are representative of extragalactic nonthermal radio sources.
In particular, the strong radio source located at ({\it l, b}) = (119\fdg38, -0\fdg84), which is close to the bright part of  Sh2-173,
has a flux density of 235 mJy at 
365 MHz and of 67.1 mJy at 1420 MHz, resulting in a spectral index $\alpha \sim -0.9$.  
The contribution of the compact sources  projected
onto the \hii\ region was subtracted in order to properly estimate the flux density of the ionized region.
The derived flux densities of
Sh2-173 are listed in Table \ref{tabflux}. The least square fitting of these values
gives a spectral index  $\alpha = 0.0 \pm 0.1$, confirming that the extended source is an \hii\ region.

\begin{table}
\centering
\begin{minipage}{84mm}
\caption{Radio continuum flux densities of Sh2-173.
\label{tabflux}}
\begin{tabular}{l c}
\hline
Frequency (MHz) & Integrated flux density (Jy)\\
\hline
325  & 1.8 $\pm$ 1.0\\
408  & 1.3 $\pm$ 0.6\\
1420 & 1.8 $\pm$ 0.3\\
2695 & 1.8 $\pm$ 0.3\\
4850 & 1.6 $\pm$ 0.3\\
\hline

\end{tabular}
\medskip

\end{minipage}
\end{table}

The parameters of the ionized gas were derived from the image at 1420
MHz. The electron density and the \hii\ mass were obtained using the
expressions of \citet{me67} for a spherical \hii\ region
of constant electron density (rms electron density $n_e$), adopting an electron temperature of 9000 K \citep{qui06}.
The derived  parameters are listed in Table~\ref{tabparam}.  

The number of UV photons necessary to keep the gas ionized was derived using $N_{Ly-c}(10^{48}\, s^{-1}) = 3.51 \times 10^{-5} \, n_e^2 (\rm cm^{-3}) \, R_s^3 (\rm pc) $. Errors in
the linear radius, in the rms electron density, and in the number of UV
photons come mainly from the distance uncertainty.

Bearing in mind that massive stars have a copious UV flux capable of
ionizing the surrounding \hi\ gas, we investigate  whether
the massive stars in Sh2-173 can provide the energy to ionize the gas.
As described in Sect. \ref{intro}, the only O-type star  among the
exciting stars is BD+60 39, being the other stars B ones.
Given that the number of ionizing photons emitted by an O-type star is significantly larger than that of a B-type star, we neglect the contribution of the UV photons emitted by the stars ALS 6145, 6150, 6155, 6156, 6157, and 6158.
 The UV photon flux corresponding to an O9V star is in the range $N_*$ (s$^{-1}$) = (1.1 - 3.6) $\times 10^{48}$ \citep*{vac96, mar05}.
By comparing $N_{*}$ with the UV photons needed to keep
the gas ionized  ($N_{Ly-c}$) listed in Table \ref{tabparam},
we conclude that the O-type star alone can maintain the \hii\ region ionized.

\begin{table}
\centering
\begin{minipage}{90mm}
\caption{Main parameters of the observed structures related to Sh2-173. \label{tabparam}}
\begin{tabular}{l c c}
\hline
Adopted distance (kpc) & \multicolumn{2}{c}{ 2.5 $\pm$ 0.5} \\
\hline
{\bf Ionized gas} & &\\
Flux density at 1420 MHz (Jy) & \multicolumn{2}{c}{$1.8 \pm 0.3$} \\
Angular radius (arcmin) & \multicolumn{2}{c}{13}\\
Linear radius (pc) &  \multicolumn{2}{c}{  9.5 $\pm$ 2.0 }\\
Ionized mass (M$_{\odot}$) &  \multicolumn{2}{c}{  480 $\pm$ 250}   \\
rms electron density $n_e$ (cm$^{-3}$) &  \multicolumn{2}{c}{6 $\pm$ 1}\\
Emission measure (pc cm$^{-6}$) &  \multicolumn{2}{c}{1000 $\pm$ 50 }\\
Used Lyman UV photons [$N_{Ly-c}$] (s$^{-1}$) & \multicolumn{2}{c}{ ( 9.3 $\pm$  4.0) $\times 10^{47}$}\\
\hline
{\bf IR emission and dust parameters} & &\\
S$_{12}$ (Jy) &    \multicolumn{2}{c}{50 $\pm$ 12}    \\
S$_{25}$ (Jy) &    \multicolumn{2}{c}{67 $\pm$ 30}     \\
S$_{60}$ (Jy) &    \multicolumn{2}{c}{590 $\pm$ 170 }    \\
S$_{100}$ (Jy)  &  \multicolumn{2}{c}{1100 $\pm$ 500 }    \\
L$_{IR}$ (L$_{\odot}$) &  \multicolumn{2}{c}{ ( 9 $\pm$  4)$\times$10$^3$   }    \\
M$_d$ (M$_{\odot}$) &   \multicolumn{2}{c}{  1.7 $\pm$  0.7  }  \\
T$_d$ (K) &  \multicolumn{2}{c}{30 $\pm $5} \\
\hline
{\bf \hi\ structure} & \\
Central coordinates & \multicolumn{2}{c}{119\fdg45, $-$0\fdg93}\\
Angular radius (arcmin) & \multicolumn{2}{c}{ 14 }\\
Linear radius (pc) & \multicolumn{2}{c}{ 10 $\pm$ 2  }\\
Velocity interval ($v_1, v_2,$ \kms) &  \multicolumn{2}{c}{$-31.1$, $-24.6$}\\
Systemic velocity (\kms) & \multicolumn{2}{c} {$-27.0 \pm 1.3$ } \\
Expansion velocity (\kms) & \multicolumn{2}{c}{5  $\pm$ 2}  \\
Velocity width ($\Delta v$, \kms) & \multicolumn{2}{c} {6$\pm$ 2}\\
Neutral atomic mass in the shell, M$_{\rm sh}$ (M$_{\odot}$) & \multicolumn{2}{c}{ 150 $\pm$ 100}\\
Missing atomic mass in the cavity,  M$_{\rm miss}$ (M$_{\odot}$) & \multicolumn{2}{c}{ 120 $\pm$ 80}\\
\hline
{\bf Molecular clouds} & Cloud A & Cloud B\\
Mean temperature (K) & 0.9$\pm$ 0.1& 1.2 $\pm$ 0.1\\
 Mean area (pc$^2$) &  80 &  20\\
Velocity interval ($v_1, v_2,$  \kms) &   $-38.6$, $-27.8$  &   $-39.4$, $-34.4$\\
Systemic velocity (\kms) &  --32.4 $\pm$  1.0  &  --36.5 $\pm$ 1.0 \\
Mean H$_2$ column density ($10^{21}$\,mol\,cm$^{-2}$) & 1.9 $\pm$ 0.6   & 1.1 $\pm$ 0.3\\
Molecular mass ($10^{3}$ M$_{\odot}$) &  5.8 $\pm$ 2.0 &  1.0 $\pm$ 0.3\\
Volume density $n_H$ (cm$^{-3}$) &  520 $\pm$ 210 &  560 $\pm$ 175 \\
\hline

\end{tabular}
\medskip
\end{minipage}
\end{table}

\subsection{The Dust}

\begin{figure*}
\includegraphics[width=18cm]{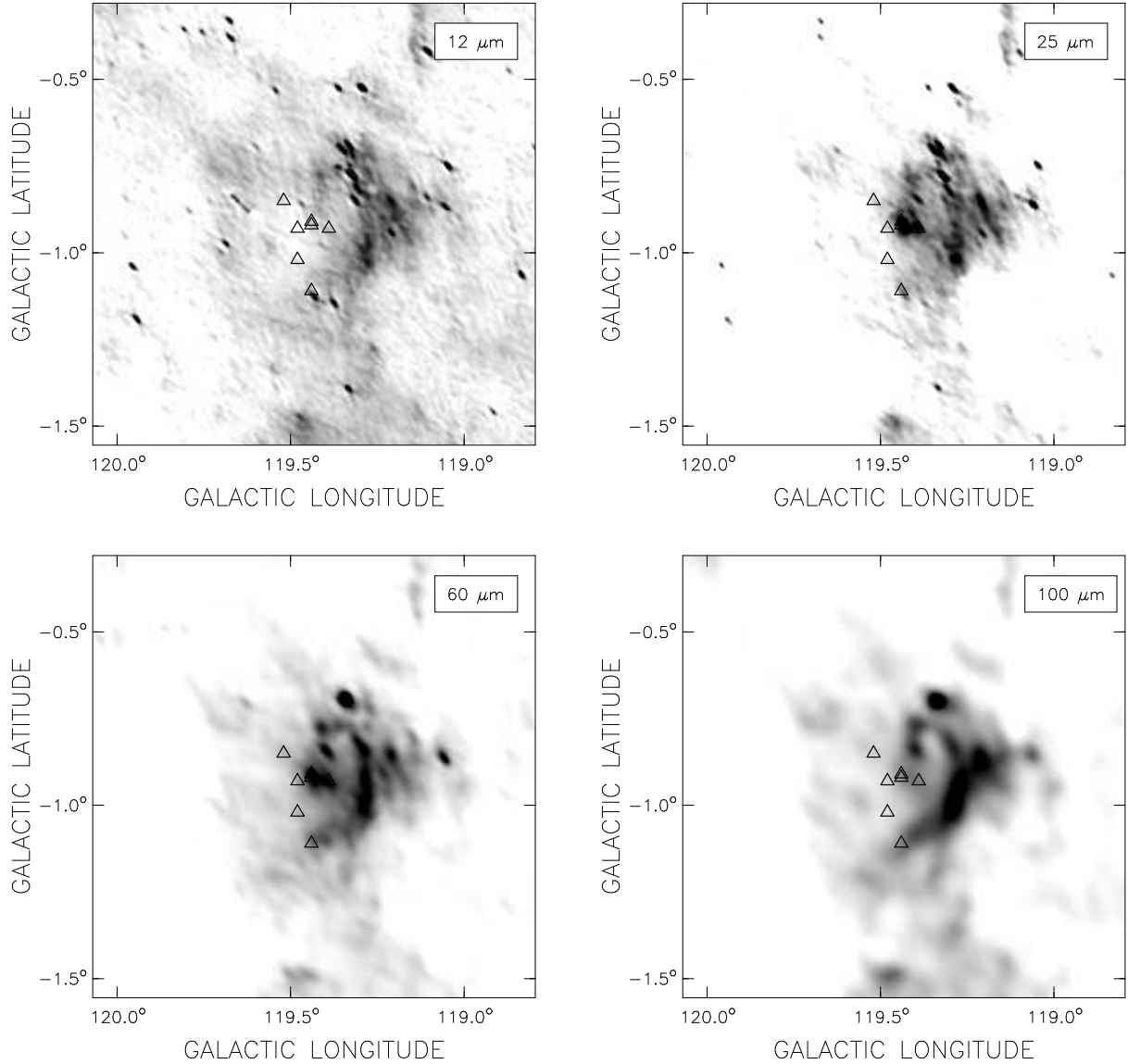}
\caption{HIRES IRAS maps of the same area as in Fig. \ref{set-continuo}. The small
  triangles indicate the position of the seven OB stars
  listed in Table \ref{tabest}. Grey-scale goes from 3 to 10 MJy\,sr$^{-1}$ (dark grey) at 12 $\mu$m, from 7 to 13  MJy\,sr$^{-1}$ at 25 $\mu$m, from 17 to 55  MJy\,sr$^{-1}$ at 60 $\mu$m, and from 70 to 160 MJy\,sr$^{-1}$ at 100 $\mu$m.}
\label{set-iras}
\end{figure*}

The HIRES IRAS data have been searched for heated dust counterpart to the ionized hydrogen region.
Figure \ref{set-iras} shows the  12, 25, 60, and 100 $\mu$m images. The image at 12 $\mu$m reveals a circular structure which exhibits less emissivity in its central part and  is brightest towards lower galactic longitudes. 
A more diffuse emission distribution is observed at 25 $\mu$m. 
The  brightest side at this wavelength coincides with the strongest radio continuum emission.
The images at  60 and 100 $\mu$m exhibit  morphological features similar to those observed at 25 $\mu$m. 
The observed IR morphology  and the position  of the OB stars  suggest that the dust is being heated by the stellar radiation.

\begin{figure*}
\includegraphics[width=18cm]{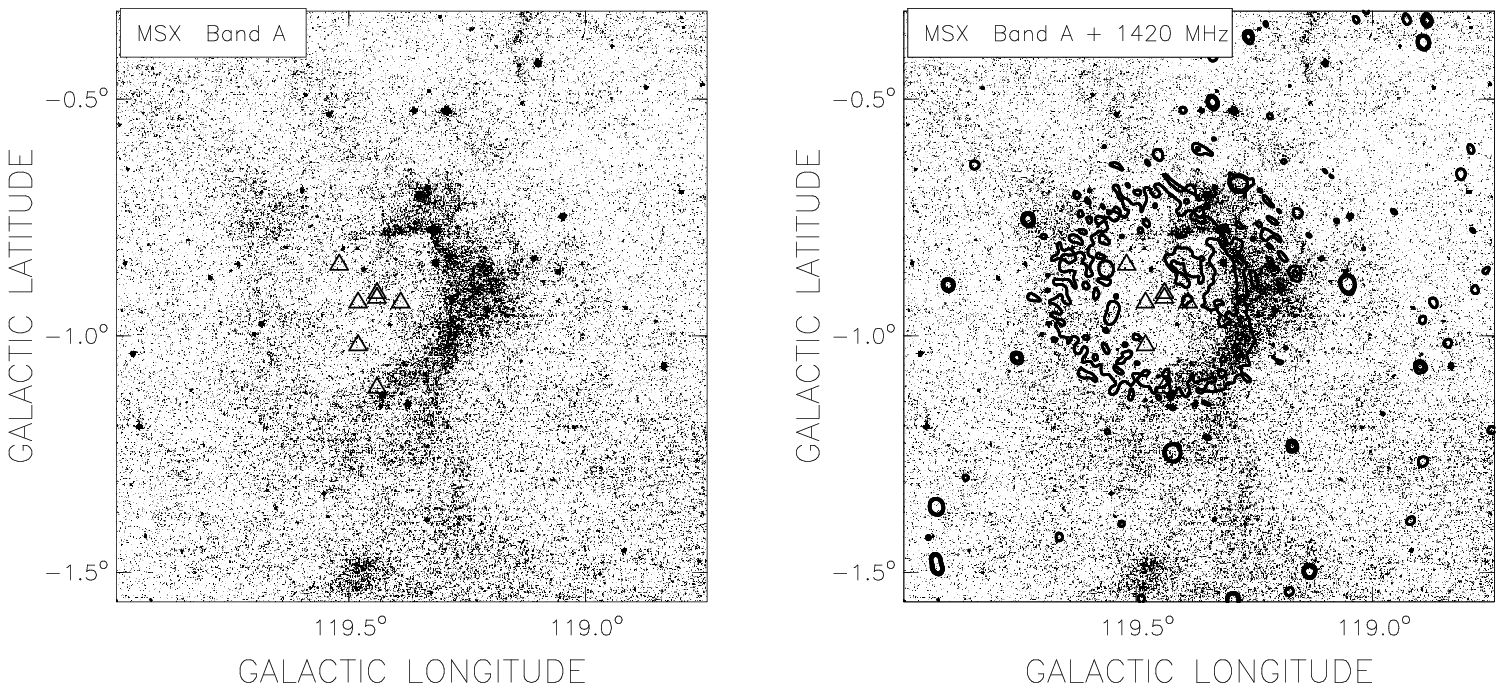}
\caption{{\it Left panel}: MSX Band A (8.3 $\mu$m) image of the area around Sh2-173.  
{\it Right panel}: Overlay of the 8.3 $\mu$m image (grey-scale) and 1420 MHz emission (contours). Contour levels are at 6 and 7 K. The small triangles indicate the position of the seven stars listed in Table \ref{tabest}.}
\label{msx}
\end{figure*}

The left panel of Fig.~\ref{msx} exhibits the mid-IR distribution at 8.3
$\mu$m (MSX Band A). The image resembles the IR distribution observed at 12 $\mu$m.
In the right panel a comparison between the IR emission at 8.3 $\mu$m ({\it grey-scale}) and the  radio continuum emission at 1420 MHz ({\it contours}) is shown. Clearly, the IR emission is mostly seen projected outside the continuum emission. 
The arc feature observed at 8.3 $\mu$m  is not detected in the other three MSX bands (named  C, D, and E).
These facts suggest that the polycyclic aromatic hydrocarbons (PAHs) could be the
main responsible for the emission detected at 8.3 $\mu$m, indicating the
existence of a photodissociated region (PDR) in the border of the \hii\
region.
The absence of 8.3 $\mu$m emission in the interior of the ionized region can be interpreted as the destruction of PAH molecules by the extreme ultraviolet radiation of the ionizing  stars.
The fact that there is emission  detected in the 25 $\mu$m IRAS image but not in the MSX E-Band   could be due to the lower sensitivity of the
MSX instruments.

 We include the main IR parameters in Table~\ref{tabparam}. 
The IR luminosity L$_{IR}$, the
interstellar dust mass  M$_d$, and the dust temperature $T_d$ were derived assuming a distance  $D$ =  2.5 $\pm$ 0.5 kpc.  
The IR luminosity was estimated following \citet{cha95} as 
$L_{IR}(L_{\odot}) = 1.58 \, S_{\rm IR}(\rm Jy)\, D^2$ (kpc), where $S_{IR}$ is the integrated flux given by $S_{IR} = 1.3(S_{12} + S_{25})+0.7(S_{25} + S_{60})+0.2(S_{60} + S_{100})$.
In order to derive the amount of radiatively heated dust, we have used the equation \citep{dra90} $M_d (M_{\odot}) = m_n\, S_{60}\ D^2 (\rm kpc)\ (B_n^{5/2} - 1)$, where $m_n = 0.30 \times 10^{-6}$ g cm$^{-2}$ Hz$^{-1}$  for the adopted value of $n$ ($n = 1.5$) and   $B_n$ is the modified Planck function, $B_n = 1.667^{3 + n} \ S_{100}/ S_{60}$.
As for the dust temperature, it was obtained as 
\citep{dra90} $T_d (K) = (95.94 / \rm ln\, B_n)$.

\subsection{The neutral hydrogen}\label{hi}

As mentioned in Section \ref{intro}, molecular emission related to Sh2-173 was found at $-34.5$ \kms\,  \citep{bli82}, 
which  coincides with the  H$\alpha$ line velocity obtained for the ionized gas, $-34.3 \pm 0.3$ \kms\,  \citep{fic90}.
Thus, although the whole
\hi\ cube was inspected looking for a structure related to the ionized
region, special attention was paid to the velocity interval $-34 \pm
15$ \kms.

\begin{figure*}
\includegraphics[]{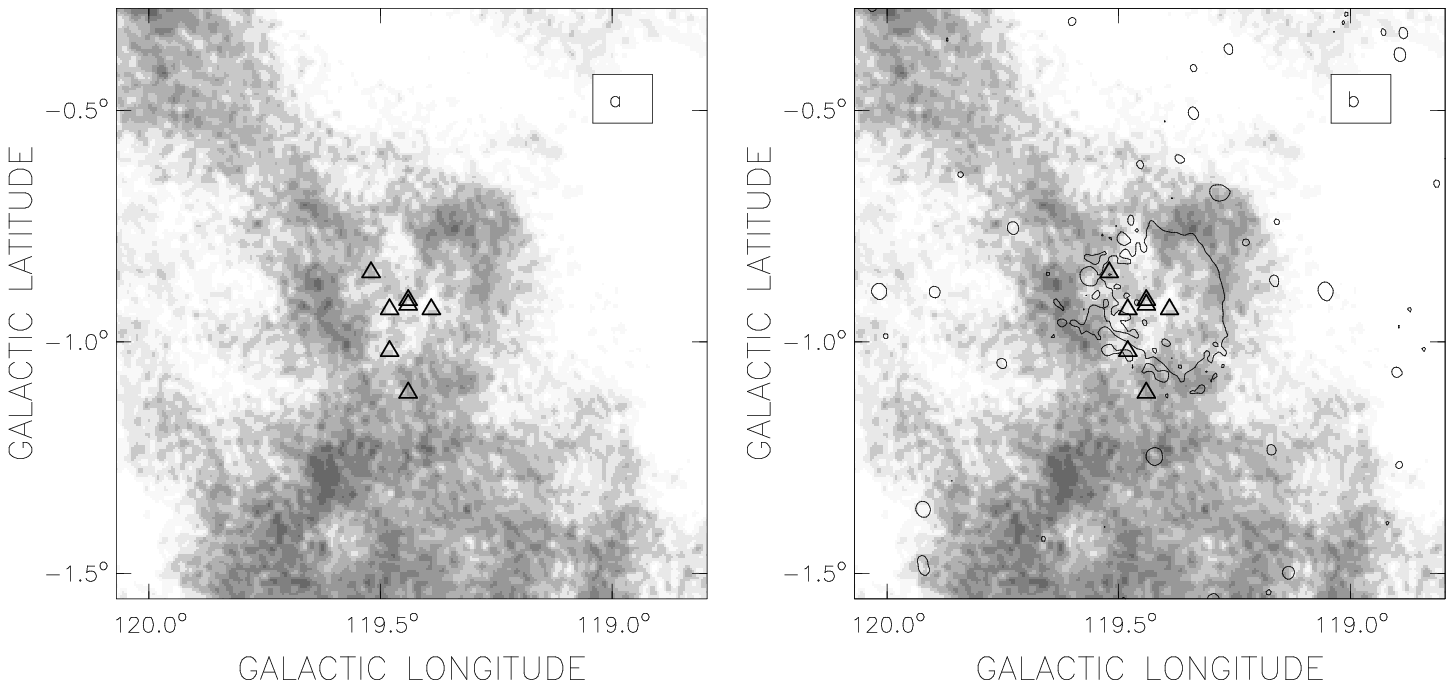}
\caption{{\it a}) Averaged \hi\ distribution in the velocity interval
  $-30.3$  to $-25.4$ \kms\ showing the \hi\ minimum related to
  Sh2-173. Grey-scale goes from 35 to 65 K. {\it b}) Overlay of the
  averaged \hi\ distribution (grey-scale) and the 6 K radio continuum
  level at 1420 MHz showing the close correspondence between the
  \hii\ region and the \hi\ cavity. The small triangles indicate the
  position of the stars listed in Table \ref{tabest}.}
\label{hi+cont}
\end{figure*}

Figure \ref{hi+cont}{\it a} shows the mean brightness temperature
image within the velocity range  $-30.3$ to $-25.4$ \kms. 
The presence of a small region of low \hi\ emission centered at ({\it
  l, b}) = (119\fdg5, $-$1\fdg0) is evident. This \hi\ void is surrounded by a region of enhanced
\hi\ emission. All of the seven stars but one (ALS\,6155) listed in Table \ref{tabest} lie inside the \hi\ minimum. ALS\,6155 
is seen projected onto the southern border of the cavity.
In Figure \ref{hi+cont}{\it b}, the 1420 MHz radio continuum image is
overlaid in contours for comparison.  The positional coincidence is
excellent, indicating that the  \hi\ feature and the
\hii\ region Sh2-173 are quite likely physically related.

A number of parameters characterizing this \hi\ structure can be
derived.
Given its angular
radius of  about 14\am\,, we calculate a physical size of
 10 $\pm$ 2 pc.  
The expansion velocity was estimated from the
velocity range spanned by the \hi\ feature as ($v_2 - v_1)/2 +
1.3$,  where $v_1$ and $v_2$ are the minimum and maximum velocities where the structure is observed, and the extra 1.3 \kms\, 
takes into account that the \hi\ caps  could be present at $v_1 - 1.3$ \kms\, and $v_2 + 1.3$ \kms.
The systemic velocity of the structure corresponds to the
velocity where the dimension of the cavity is largest.  The mass of
the \hi\ ring can be calculated under the assumption of optically thin
\hi\ emission. Following the procedure described by \citet{pin98}, the
\hi\ mass is given by $\rm M_{\hi }(\rm M_{\odot}) = 1.3\, \times\,
10^{-3}\, \rm D^2_{\rm kpc}\,$ $\Delta v\, \Omega_{am^2}\, \Delta \rm
T_{\rm B}$ , where $\rm D^2_{\rm kpc}$ is the distance to the
\hi\ structure and $\Delta v$  the velocity width  over which the \hi\ cavity is detected, in \kms. 

The
meaning of $\Omega_{am^2}$ and $\Delta \rm T_{\rm B}$ depends on
whether we want to find the missing mass in the cavity or the excess
mass in the shell. In the first case $\Delta \rm T_{\rm B}$ is the
brightness temperature difference between the void and the background,
and $\Omega_{am^2}$ is the cavity solid angle in arcmin$^2$. In the
second case $\Delta \rm T_{\rm B}$ is the brightness temperature
difference between the shell and the background, and $\Omega_{am^2}$ is
the solid angle of the shell. 
All the derived parameters are  listed in Table \ref{tabparam}.

\subsection{CO emission distribution}\label{co}

\begin{figure*}
\includegraphics[]{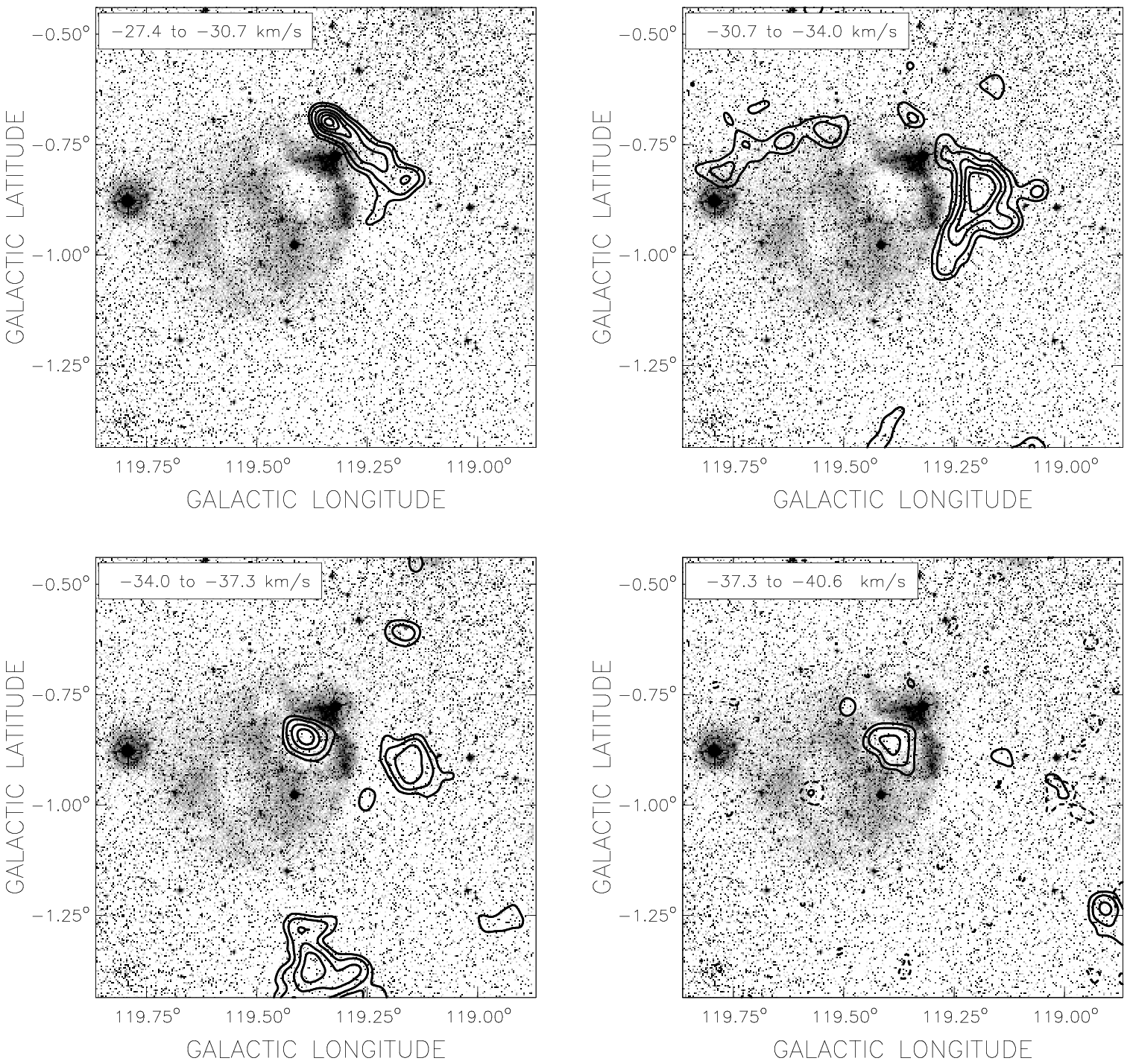}
\caption{\small{Ionized and molecular gas in the environs of
    Sh2-173}. The images display overlays of the CO emission distribution in the velocity
    range from  --27.4 to --40.6 \kms\, ({\it contours}) and the DSS-R
    optical image ({\it grey-scale}). Contour levels are 0.5 K
    ($\equiv$ 5 $\sigma$), and from 1 to 5 K in steps of 1 K. Each
    panel shows the average of four consecutive CO maps. The dotted contour shown in the last velocity interval corresponds to the 0.6 K level of the image at --40.2 \kms. }
\label{setco}
\end{figure*}

\begin{figure*}
\includegraphics[]{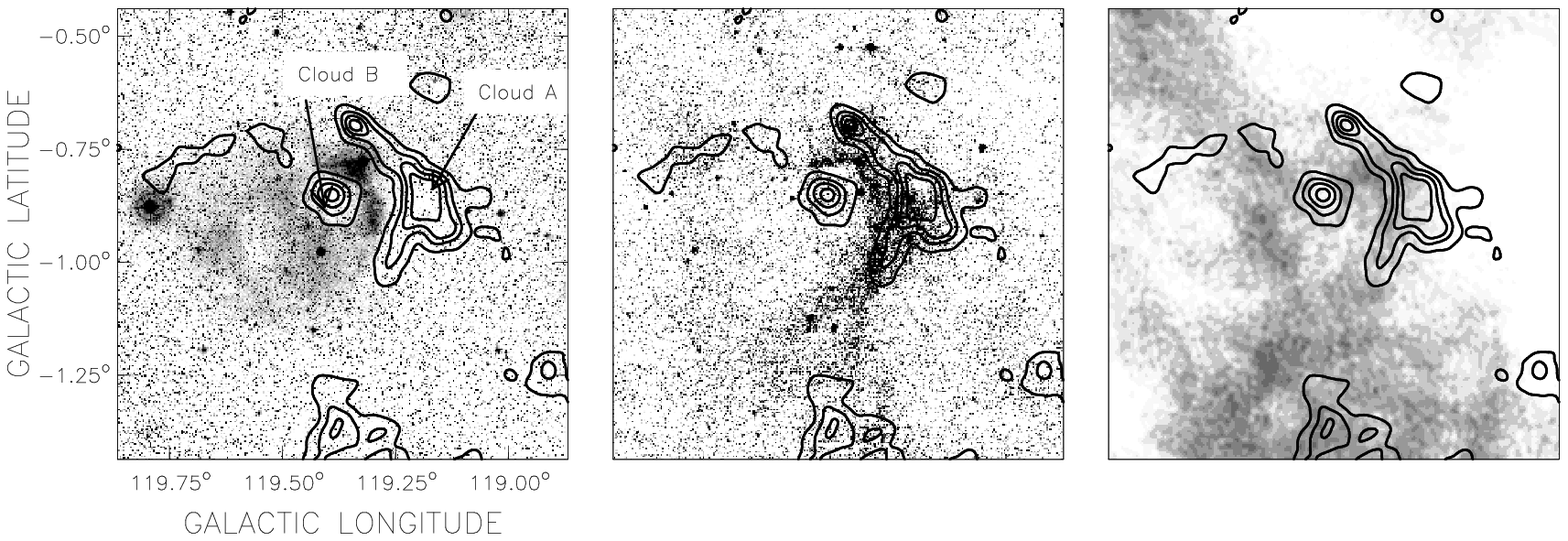}
\caption{\small Dust and gas in the environs of Sh2-173.  In the three panels the contours show the 
CO  emission distribution within the velocity range from
   --27.4 to --39.0 \kms. Contour levels are from 0.2
  to 1.8 in steps of 0.4 K.
 {\it Left
    panel}: The CO contours are superimposed on the 
 DSS2-R optical image ({\it grey-scale}).
{\it Middle panel}:  The CO contours are superimposed on the 
 MSX Band A image.  {\it Right panel}: The CO contours are superimposed on the averaged \hi\ distribution.}
\label{co-comp}
\end{figure*}

Figure \ref{setco} shows the $ ^{12}$CO(1--0) emission distribution within the velocity
interval from  --27.4 to  --40.6 \kms\, in steps of 3.3 \kms\, ({\it contours})
 overimposed onto the DSS2-R optical image ({\it grey-scale)}. Each panel is the
average of four consecutive CO maps.

In the velocity interval from  --27.4 to --30.7 \kms\,, the CO emission partially
borders the brightest optical emission arc  at ({\it l, b}) $\approx$ (119\fdg25,
--0\fdg75). In the second velocity interval, from  --30.7 to  --34.0 \kms, the CO structure extends 
towards the south increasing in intensity at
lower galactic longitudes. Low molecular arc-shaped emission is distinguished
surrounding the optical nebulosity at ({\it l, b}) $\approx$ (119\fdg6, --0\fdg75).
The molecular emission distribution in the velocity range from  --34.0 to
--37.3 \kms\, changes in comparison with the previous images. Several different CO features
are observed, being the dominant one  a CO clump which spatially
coincides with the area lacking optical emission at ({\it l, b}) $\approx$ (119\fdg4,
--0\fdg85). The fact that radio continuum emission is detected in this region
(see Fig. \ref{set-continuo}b), reinforces the possibility that this molecular clump is the
responsible for the observed optical absorption feature.
In the last velocity interval, from  --37.3 to --40.6 \kms, the CO structures seen from  --27.4 to  --34.0
\kms\, disappear and most of the CO emission appears projected onto the
optical absorption features situated in the central part of the \hii\ region.
In two of the velocity channels corresponding to this velocity interval,
hardly detected (3 rms) molecular emission is seen projected onto ({\it l, b}) $\approx$
(119\fdg57, --1\fdg0), in coincidence with the other area  where optical emission is absent.
This CO structure is  indicated by a dotted contour in Fig. \ref{setco} .

In Fig. \ref{co-comp} the averaged CO
emission distribution in the velocity interval from  --27.4 to --39.0 \kms\,  in comparison with the
ionized gas, the dust, and the \hi\, distributions in the environs of Sh2-173 is shown. The left panel of Fig. \ref{co-comp} shows
the averaged CO emission distribution ({\it contours})
   overimposed onto the DSS2-R optical image ({\it grey-scale}). There are two
main clouds of molecular gas probably related to Sh2-173. Towards lower galactic longitudes, 
molecular gas partially encircles the \hii\ region (cloud A from here on). The dense molecular gas in this region has probably slowed
down the ionization front and the expansion of the \hii\ region in this
direction. On the other hand, the area near ({\it l, b}) = (119\fdg4, --0\fdg85),
which depicts relatively strong radio emission and is almost free of optical
emission, coincides with a cloudlet of molecular material detected within the
velocity range  from  --34.0 to --40.6 \kms\, (Cloud B in Fig. \ref{co-comp}). This molecular gas
and the associated dust should be in front of the ionized material to originate the observed optical absorption. The velocity of 
cloud B ($\sim -38$ \kms) is compatible with this material being pushed towards the observer.
 The difference in radial velocity between cloud A and cloud B allows us to get a rough estimation of the expansion velocity of the molecular gas, which results to be $4 \pm 2$ \kms.
The  middle panel of Fig.\ref{co-comp} shows  an overlay
of the image at \hbox{8.3$\,\umu$m} ({\it grey-scale}) and the CO emission
distribution ({\it contours}). The mid-infrared and molecular emissions present
an excellent morphological correspondence. The molecular gas encircles the bright MSX
rim. A fainter CO emission  feature is observed towards the infrared emission detected at ({\it l, b})
$\approx$ (119\fdg6, --0\fdg75) at 8.3 $\,\mu$m.
 The right panel of Fig. \ref{co-comp} shows a comparison between the CO and \hi\ emission distributions, showing that the CO structure 
(cloud A) is located slightly further from the ionized gas as compared with the \hi\ shell.

Summing up, the comparison between the spatial distribution of the ionized and
 neutral gas, together with the dust, reveals a remarkable  morphological correspondence. 
The emission of the ionized gas is partially encircled by an arc
shaped feature of mid-infrared emission and an outermost feature of
molecular gas. This stratified distribution suggests that the ionization front
is bordered by a PDR,  detected as the bright rim at 8.3 $\,\mu$m.

Table~\ref{tabparam} summarizes the physical parameters of the two
molecular clouds shown in Fig. \ref{co-comp} (named cloud A and cloud B) as
derived from the $^{12}$CO data.  The masses were estimated by
integrating the CO line intensity as $W{\rm_{CO} =\int{T(CO)dv}}$, where
T(CO) is the average temperature of the molecular gas over the
velocity interval  where each cloud is observed.
To calculate the H$_2$ column density,
the  relationship $X=N({\rm H_2})/W_{{\rm CO}}$ of $1.9 \times 10^{20}\ {\rm
cm^{-2} (K\ km s ^{-1})^{-1}}$ \citep*{gre90,dig95} was considered.  The molecular mass was derived from
$M[M_\odot]=4.2 \times 10^{-20}N(H_2)D^2 A$, where $D$ is the distance in pc and $A$  is the  solid angle in  steradians.  Assuming a
spherical geometry for both clouds, the volume density can be
estimated as $n=9.79 \times M[M_\odot]/R^3$, where $R$ is the
radius of the molecular cloud in pc.

In order to obtain an estimate of the visual absorption $A_V$ produced by the gas in the direction of Sh2-173,
 we have used both the \hi\ and CO data cubes.
The  column densities $N(\hi)$ and $N(H_2)$ were obtained by integrating the data cubes in the velocity interval from 0 to $-40$ \kms\,  
 in a small box area ($\sim 2\arcmin$)  over cloud A.
The total hydrogen column density was obtained as $N_H = N(\hi) + 2\, N(H_2)$. 
Using the standard conversion factor, the total visual extinction was estimated as $A_V= 5.2 \times 10^{-22}\, N_H$ \citep*{boh78}, 
which in this case yields $A_V = 4 \pm 1$ mag.
 For comparison, the visual extinction towards the OB stars listed in Table \ref{tabest}, where no molecular emission is observed, is in the range from 1.3 to 2 mag \citep{rus07}.

\section{Stellar formation activity}\label{star-formation}

The multiwavelength study of Sh2-173 reveals that it is a classical expanding \hii\ region  containing gas and dust, partially encircled 
by a patchy molecular emission. The interface between the ionized and the molecular gas is clearly evident with the detection of a PDR. 
The observed morphology leads to the question of whether the expansion of the ionized region could have induced the formation of a new 
stellar generation in the neighboring molecular envelope.
In order to analyze this possibility,  we have made a systematic search for young stellar object (YSO) candidates in the area.

To look for primary tracers of stellar formation activity, we used the MSX6C Infrared Point Source Catalogue \citep{ega03} in Bands A 
(\hbox{8.3 $\,\umu$m}), C (\hbox{12.1 $\,\umu$m}), D (\hbox{14.7$\,\umu$m}), and E (\hbox{21.3$\,\umu$m}); the 2MASS All-Sky Point Source 
Catalogue \citep{cut03} in bands $J$ (1.25 $\,\umu$m), $H$ (1.65 $\,\umu$m), and $K$ (2.17 $\,\umu$m); and the IRAS Point Source 
Catalogue \footnote{1986 IRAS catalogue of Point Sources, Version 2.0 (II/125)}. 

YSO candidates were searched in a 1\degr\, $\times$  1\degr\, box area centered at \hbox{({\it l, b}) = (119\fdg36, --0\fdg9 )}.
A total of 30 IRAS point sources were found projected onto the analyzed region. \citet*{jun92}'s conditions for YSOs are: S$_{100}$ $\ge$
 20 Jy, \hbox{1.2 $\leq$ $\frac{S_{100}}{S_{60}}$ $\leq$ 6.0}, $\frac{S_{60}}{S_{25}}$ $\ge$ 1, and Q$_{60}$+Q$_{100}$ $\ge$ 4, 
where S$_{\lambda}$ and Q${_\lambda}$ are the flux density and the quality of the IRAS flux in each of the observed bands, respectively. 
Only 2 of the 30 observed IRAS sources fulfill these conditions and may be classified as protostellar candidates.

The MSX sources were classified based on the \citet{lum02}'s criteria. The selection of the sources is made taking into account their loci 
in the (F$_{21}$/F$_{8}$, F$_{14}$/F$_{12}$) diagram, where F$_{\lambda}$ denotes the flux in each band. According to these criteria, 
massive young stellar object (MYSO) candidates have \hbox{F$_{21}$/F$_{8}$ $\ga$ 2} and \hbox{F$_{14}$/$F_{12}$ $\ga$ 1}, while 
compact H{\sc ii} regions (C\hii)~present \hbox{F$_{21}$/F$_{8}$ $\ga$ 2} and \hbox{F$_{14}$/$F_{12}$ $<$ 1}. Evolved stars occupy 
the region where \hbox{F$_{21}$/F$_{8}$ $\leq$ 2} and \hbox{F$_{14}$/F$_{12}$ $\leq$ 1}. 
We have found 40 MSX sources projected onto the area, however neither of them satisfy the criteria to be classified as a MYSO nor a C\hii.

 To look for tracers of stellar formation
activity in the 2MASS catalogue, we have adopted the criteria developed by \citet{ker08}.
Based on a sample of
YSOs having low ($<$ 2 M$_{\odot}$) and intermediate (2 $<$ M$_{\odot}$ $<$ 9) masses
extracted from \citet{Ken95} (T Tauri) and \citet*{the94} (Herbig/AeBe), respectively, \citet{ker08} inferred a colour criteria 
to analyze the presence of stellar formation activity in the surroundings of the \hii\ region  KR\,140.
They  scaled the data of the sample  to the distance of 
 KR\,140 (2.3 kpc) and the corresponding visual extinction (A$_{\rm v}$ = 5.5 mag).
Using the 2MASS Point Source Catalogue, and taking into account  different photometric qualities and the adopted colour criteria, 
these authors divided the YSO candidates in four groups, named P1, P1+, P2, and P3.
P1 and P1+ sources have the best photometric quality (ph\_qual  = A, B, C,
or D). P1 sources lie below the reddening vector associated with an O6 V star, while P1+ sources lie in the overlapping region of T Tauri and main sequence stars.
The P2 group contains sources which have not been detected in the
J-band (ph\_qual  = U in this band). 
In this case, the adopted  J-magnitude is a lower limit and consequently the corresponding value of (J-H)  is a lower limit as well.
Thus, the actual position of the P2 sources in the (J-H)-axes is likely to be towards higher (J-H) values.
Finally, the sources belonging to the P3 group have J and H magnitudes that are lower limits, so their colour (J-H) can not be considered.
These sources can not be included in the CC diagram, and
their real position on the CM diagram may be  towards higher values of (H-K).
To analyze the presence of 2MASS  YSO candidates in the surroundings of Sh2-173 we have adopted the Kerton et al.'s criteria
scaled to the distance adopted for Sh2-173 (2.5 kpc) and the visual absorption in this direction 
(A$_{\rm v}$ = 4.0 mag). The modified criteria 
are listed in Table \ref{criteria}.

\begin{table}
\centering
\caption{Selection criteria for 2MASS YSO candidates. \label{criteria}}
\begin{tabular}{lc}
\hline
Classification & Colour criteria\\
\hline
P1& ($J-H$) $>$ 0.835\\
   & ($J-H$) $<$ 1.7  ($H-K$) -- 0.087\\
   \hline
P1+& 1.135 $<$ ($J-H$) $<$ 1.435\\
    & ($J-H$) $>$ 1.7  ($H-K$) -- 0.087\\
     & ($J-H$) $<$ 1.7  ($H-K$) + 0.3685\\
    & $K$ $>$ 14.5\\
    \hline
P2 & ($J-H$)$_{L}$ $<$ 1.7 ($H-K$) -- 0.087\\
& ($H-K$) $>$ 0.91\\
\hline
P3&($H-K$)$_{L}$   $>$ 0.91\\
\hline
\hline
\end{tabular}\\
The underscript "$L$" indicates a lower limit.
\end{table}

\begin{figure*}
\centering
\includegraphics[angle=0,width=0.7\textwidth]{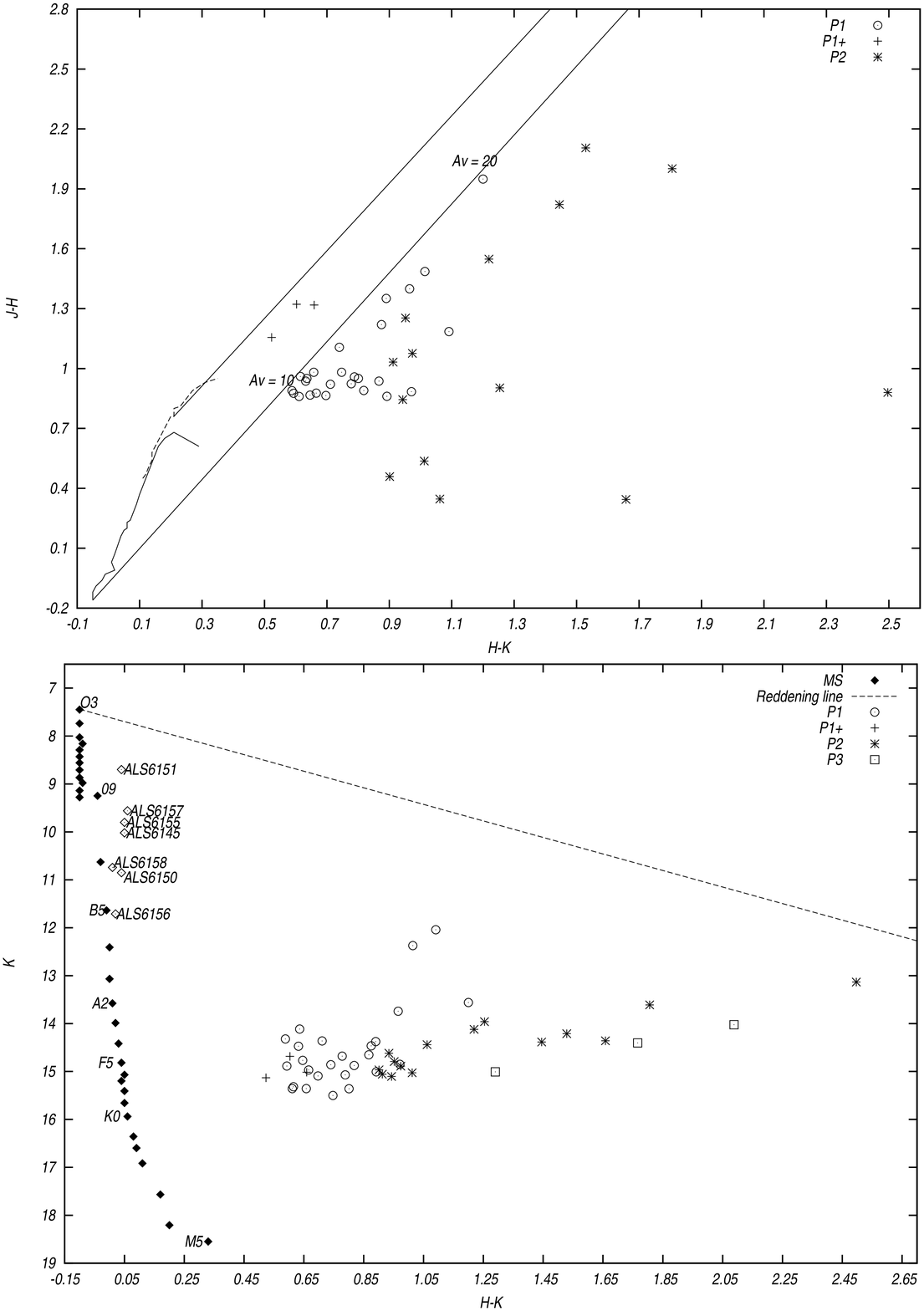}
\caption{\small{ CC and CM diagrams of the 2MASS YSO candidates  projected onto clouds A and B. P1, P1+, P2, and P3 sources are indicated by circles, crosses, asterisks, and squares, respectively. {\it Upper panel}: $(H-K, J-H)$ diagram. The positions of the derredened early-type main sequence and giant stars are shown. The reddening curves for MO III stars (upper line) and O9 V stars (lower line) are indicated. The colours $(J - H)$ corresponding to P2 sources are lower limits. {\it Lower panel}:  $(H-K, K)$ diagram.  The position of the main sequence  \citep{tok00, dri00, mar06} with A$_{\rm v} = 4$ mag at a distance of 2.5 kpc is indicated by filled diamonds. The locations of the OB stars related to Sh2-173 are indicated by their corresponding names. The reddening curve for an O3 star is shown with a dashed line. The colours $(H - K)$ corresponding to P3 sources are lower limits.\label{cc-cm}}}
\end{figure*}

A total of 23988, 135, and 134 sources from the 2MASS catalogue were examined in order to find {\it P1} or {\it P1+}, {\it P2}, and {\it P3} candidates, respectively. 
From the 23988 sources, only 81 and 10  were classified as {\it P1} and {\it P1 +}, respectively. 
Out of the 135 candidates, 19 were identified as {\it P2} sources, while only 5 out of the 134 candidates  were classified as {\it P3} sources.
Thus, a total of 115 2MASS sources were found to be YSO candidates. 
However, 2 of these sources  had to be excluded from the sample since they were identified as part of the galactic contamination in the 2MASS catalogue (i.e. galcontam=2). 
Figure \ref{cc-cm} shows the CC (top panel) and CM (bottom panel) diagrams for  the YSO candidates projected onto the molecular gas associated with Sh2-173. The CM diagram also shows the position of the Sh2-173's exciting stars. The reddening vectors for an early type (O9 V) and a late-type (M0 III) stars \citep{tok00} are represented by two parallel lines in the CC diagram using extinction values from \citet{rie85}.  For the location  of the main sequence we have used for the O-stars the values given by \citet{mar06}, and for the  late-type stars (B to M) the ones given by \citet{tok00} and \citet{dri00}. The main sequence obtained for a distance of 2.5 kpc  is indicated by  filled diamonds in the CM diagram. The location of the YSO candidates in the CC diagram is in concordance with the numerical models developed by \citet{rob06}.

\begin{figure}
\includegraphics[width= 90mm]{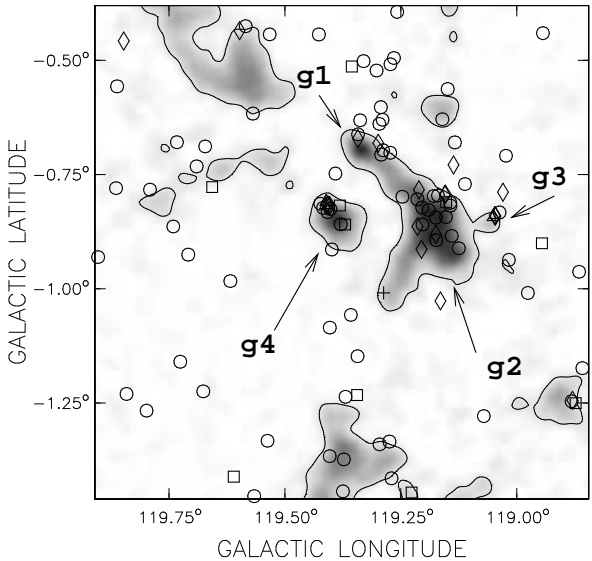}
\caption{Infrared point sources classified as YSO candidates overlaid onto the CO distribution averaged between the velocity range from  --27.4 to --49.8 \kms. IRAS, P1, P1+, P2, and P3 sources are indicated as crosses, circles, squares, diamonds, and triangles, respectively. The locations of the four groups, g1, g2, g3, and g4, described in the text are indicated. }
\label{martin}
\end{figure}

Figure \ref{martin} shows the location of the YSO candidates with respect to the  $^{12}$CO emission distribution in the velocity range from  --27.4 to --49.8 \kms. 
IRAS, P1, P1+, P2, and P3 sources are indicated as crosses, squares, circles, diamonds, and plus signs, respectively. 
As can be inferred from the figure, 65 of the 115 YSO candidates  appear projected onto molecular gas, from which 46  are located onto molecular clouds probably related to Sh2-173 (clouds A and B).
There is a clear concentration of sources towards the  brightest parts of the CO emission associated with Sh2-173.
Based on their relative location, we have separated them into four groups: g1, g2, g3, and g4 (see Fig. \ref{martin}). The groups g1 and g2 contain 8 and 20 2MASS sources, respectively, and are located in the periphery of the optical nebula, onto the PDR. The group g3 is conformed by sources located towards lower galactic longitudes, onto cloud A. 
The sources  of the group g4 are located onto cloud B.

The main data of the IRAS and 2MASS sources that are seen projected onto molecular gas  are listed in  Table \ref{ysos}. 
It is worth  mentioning that an inspection of the CO data cube reveals that there is only one velocity component in the direction of Sh2-173. This fact strongly suggests that the YSO candidates suffer from extinction due to the molecular material related to Sh2-173.

To sum up, based on different infrared colour criteria and on the location of the molecular emission, we have identified 46  YSO candidates probably associated with Sh2-173.

The presence of a PDR is a clear proof of the interaction between the  \hii\ region and the molecular gas, since it  is a region directly affected by the expansion of the ionized gas and by the UV photons of the exciting stars. 
Thus, the relative location of the YSO candidates with respect to the PDR may be considered as an indication of the role that the \hii\ region is playing in the formation of the new stars. 
The probability for a source of being triggered by the action of the \hii\ region is higher when it is  located inside the region of direct influence of the \hii\ region \citep[and references therein]{ker08}.

Figure \ref{martin} shows that there are two groups of sources, named g1 and g2, located adjacent to the PDR, while 
the group g3 is located further, towards lower galactic longitudes.
The possibility for the group g4 of being located near or onto a PDR can not be discarted. A face-on PDR, which could not be detected, may be present between the ionized region and the molegular gas.

\begin{table*}
\small
\caption{ YSO candidates projected onto the molecular clouds A and B. \label{ysos}}
\begin{tabular}{c c c l r r r r c c}
\hline
\multicolumn{9}{c}{{\bf IRAS sources}} \\
\hline
 $\#$ &  Designation &  {(\it l, b})&     \multicolumn{4}{c}{Fluxes [Jy]}  & $L_{IRAS}$[L$_{\odot}$] & Group \\
& & & 12 $\mu$m & 25 $\mu$m & 60 $\mu$m & 100 $\mu$m & For D =  2.5 kpc & \\
\hline
1&00158+6128&119\fdg057 --0\fdg87& 0.92& 2.29& 8.78&23.29 & $\sim$  180 & g3 \\
2&00179+6121&119\fdg282 --1\fdg02 & 0.32& 0.3& 5.89&23.89 & $\sim$  110 &  - \\
\hline
\multicolumn{9}{c}{{\bf 2MASS sources}} \\  \hline
$\#$ & Designation  & ({\it l, b}) &     $J$ & $H$ & $K$ &  $(H-K)$&$(J-H)$& Group \\
 \hline
 \multicolumn{9}{c}{{\bf P1}} \\
\hline
3&00183077+6145567  &   119\fdg047  --0\fdg857& 14.317 & 13.132 & 12.041 & 1.091 & 1.185 &g3\\
4&00211376+6146520  &   119\fdg368   --0\fdg881& 16.892 & 15.932 & 15.317 & 0.615 & 0.96 &g4\\
5&00193852+6143342  &   119\fdg175   --0\fdg913 & 16.816 & 15.858 & 15.07 & 0.788 & 0.958 &g2\\
6&00194881+6147427  &   119\fdg204   --0\fdg847& 15.799 & 14.91 & 14.322 & 0.588 & 0.889 &g2\\
7&00202395+6154406  &   119\fdg286   --0\fdg740 & 17.227 & 16.246 & 15.499 & 0.747 & 0.981 &g1\\
8&00201448+6154455  &   119\fdg268   --0\fdg736 & 16.76 & 15.899 & 15.007 & 0.892 & 0.861 &g1\\
9&00212578+6143577  &   119\fdg386   --0\fdg932 & 16.827 & 15.967 & 15.356 & 0.611 & 0.86 &g4\\
10&00191976+6147310  &   119\fdg146   --0\fdg843 & 16.041 & 15.104 & 14.472 & 0.632 & 0.937 &g2\\
11&00212969+6149057  &   119\fdg404   --0\fdg848 & 17.11 & 16.16 & 15.359 & 0.801 & 0.95 &g4\\
12&00193126+6148555  &   119\fdg172   --0\fdg823 & 16.706 & 15.822 & 14.851 & 0.971 & 0.884 &g2\\
13&00192242+6143509  &   119\fdg144   --0\fdg904 & 16.382 & 15.458 & 14.68 & 0.778 & 0.924 &g2\\
14&00191965+6147518  &   119\fdg147   --0\fdg837 & 16.709 & 14.76 & 13.56 & 1.2 & 1.949 &g2\\
15&00202211+6158243  &   119\fdg290  --0\fdg678 & 16.455 & 15.518 & 14.651 & 0.867 & 0.937 &g1\\
16&00193625+6146178  &   119\fdg176   --0\fdg867 & 16.28 & 15.413 & 14.767 & 0.646 & 0.867 &g2\\
17&00204518+6157266  &   119\fdg333  --0\fdg699 & 14.873 & 13.387 & 12.373 & 1.014 & 1.486 &g1\\
18&00195258+6148460  &   119\fdg213   --0\fdg830 & 16.51 & 15.633 & 14.967 & 0.666 & 0.877 &g2\\
19&00212565+6148318  &   119\fdg395   --0\fdg856 & 16.559 & 15.339 & 14.464 & 0.875 & 1.22 &g4\\
20&00195062+6145397  &   119\fdg203   --0\fdg881 & 15.993 & 15.072 & 14.361 & 0.711 & 0.921 &g2\\
21&00194259+6147191  &   119\fdg191  --0\fdg852 & 16.584 & 15.694 & 14.876 & 0.818 & 0.89 &g2\\
22&00192579+6146125  &   119\fdg156   --0\fdg866 & 15.701 & 14.751 & 14.115 & 0.636 & 0.95 &g2\\
23&00213187+6149382  &   119\fdg409   --0\fdg839 & 16.106 & 14.707 & 13.742 & 0.965 & 1.399 &g4\\
24&00212247+6148506  &   119\fdg389   --0\fdg850 & 16.616 & 15.265 & 14.375 & 0.89 & 1.351 &g4\\
25&00204129+6159065  &   119\fdg329   --0\fdg671 & 16.997 & 16.016 & 15.358 & 0.658 & 0.981 &g1\\
26&00193569+6148554  &   119\fdg180  --0\fdg824 & 16.703 & 15.597 & 14.857 & 0.74 & 1.106 &g2\\
27&00191679+6142153  &   119\fdg130  --0\fdg929 & 16.654 & 15.789 & 15.092 & 0.697 & 0.865 &g2\\
28&00202147+6155100  &   119\fdg282  --0\fdg731 & 16.356 & 15.48 & 14.887 & 0.593 & 0.876 &g1\\
\hline
\multicolumn{9}{c}{{\bf P1+}} \\
\hline
29&00210933+6146474  &   119\fdg359 --0\fdg881 & 16.811 & 15.655 & 15.132 & 0.523 & 1.156 &g4\\
30&00192372+6147592  &   119\fdg155 --0\fdg836 & 16.992 & 15.674 & 15.015 & 0.659 & 1.318 &g2\\
31&00211236+6149088  &   119\fdg370  --0\fdg843 & 16.609 & 15.287 & 14.684 & 0.603 & 1.322 &g4\\
\hline
\multicolumn{9}{c}{{\bf P2}} \\
\hline
32&00182469+6148202  &   119\fdg041  --0\fdg816 & $>$16.575& 16.038& 15.026 & 1.012 & $>$0.537 &g3\\
33&00192419+6148555  &   119\fdg158   --0\fdg821 & $>$17.416& 15.415 & 13.609 & 1.805 & $>$2.001 &g2\\
34&00193839+6144163  &   119\fdg176   --0\fdg901 & $>$16.996& 15.743 & 14.791 & 0.952 & $>$1.253 &g2\\
35&00212409+6149210  &   119\fdg393  --0\fdg842 & $>$16.361& 16.017 & 14.359 & 1.657 & $>$0.344 &g4\\
36&00202552+6156054  &   119\fdg292  --0\fdg717 & $>$16.886 & 15.338 & 14.119& 1.219 & $>$1.549 &g1\\
37&00192337+6148577  &   119\fdg156  --0\fdg820 & $>$17.844& 15.739 & 14.21 & 1.529 & $>$2.106 &g2\\
38&00212265+6149170  &   119\fdg390  --0\fdg843 & $>$16.938 & 15.862 & 14.888 & 0.973 & $>$1.077 &g4\\
39&00204593+6156560  &   119\fdg334   --0\fdg708 & $>$16.118 & 15.215& 13.961& 1.255 & $>$0.902 &g1\\
40&00212591+6149307  &   119\fdg397  --0\fdg840& $>$14.951& 15.553 & 14.619 & 0.934 & $>$0.601&g4\\
41&00195601+6145301  &   119\fdg213   --0\fdg885 & $>$16.325 & 15.866 & 14.965 & 0.901 & $>$0.458 &g2\\
42&00195468+6142421  &   119\fdg205  --0\fdg931 & $>$17.000& 15.968 & 15.0559& 0.913 & $>$1.032 &g2\\
43&00194992+6149530  &   119\fdg210  --0\fdg811 & $>$15.847& 15.501 & 14.439 & 1.063 & $>$0.346 &g2\\
\hline
\multicolumn{9}{c}{{\bf P3}} \\
\hline
44&00183752+6145299  &   119\fdg060   --0\fdg866 & $>$18.312& $>$16.298 & 15.008 & $>$1.29 & -&g3\\
45&00212628+6149244  &   119\fdg397   --0\fdg842 & $>$18.274 & $>$16.110 & 14.022& $>$2.087 & - &g4\\
46&00212574+6149479  &   119\fdg397   --0\fdg835 & $>$18.1310& $>$16.170 & 14.404& $>$1.766 & - &g4\\
\hline
\end{tabular}
\end{table*}

\section{The Big picture}

\begin{figure*}
\includegraphics[width=15cm]{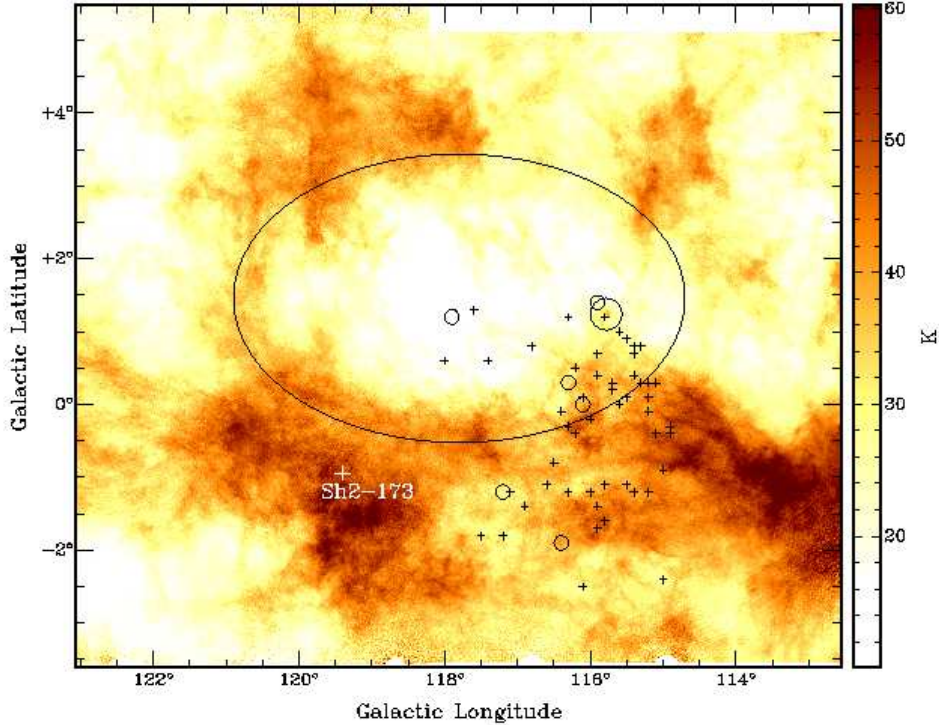}
\caption{Averaged \hi\ emission distribution between the velocity range from --50 to --20 \kms. 
The circles and crosses indicate the position of the stars belonging to Cas\,OB5 according to \citet{gar92}. The small circles correspond to O-type stars, while crosses mark the location of B stars. The large circle  indicates the position of WR\,159. The ellipse delineates the location of the \hi\ shell. }
\label{gran-shell-hi}
\end{figure*}

\begin{figure*}
\includegraphics[width=15cm]{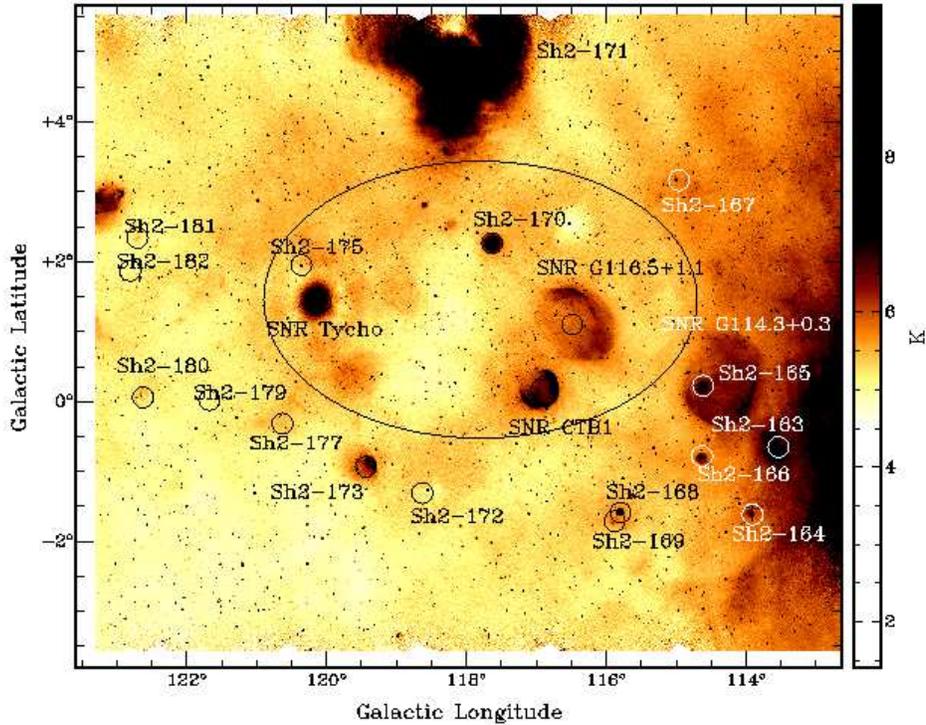}
\caption{1420 MHz radio continuum image from the CGPS survey. The ellipse delineates the \hi\ shell shown in Fig.\ref{gran-shell-hi}. The location of the Sharpless \hii\ regions and the SNRs present in the area are indicated by circles.}
\label{gran-shell-1420}
\end{figure*}

\begin{figure*}
\includegraphics[width=15cm]{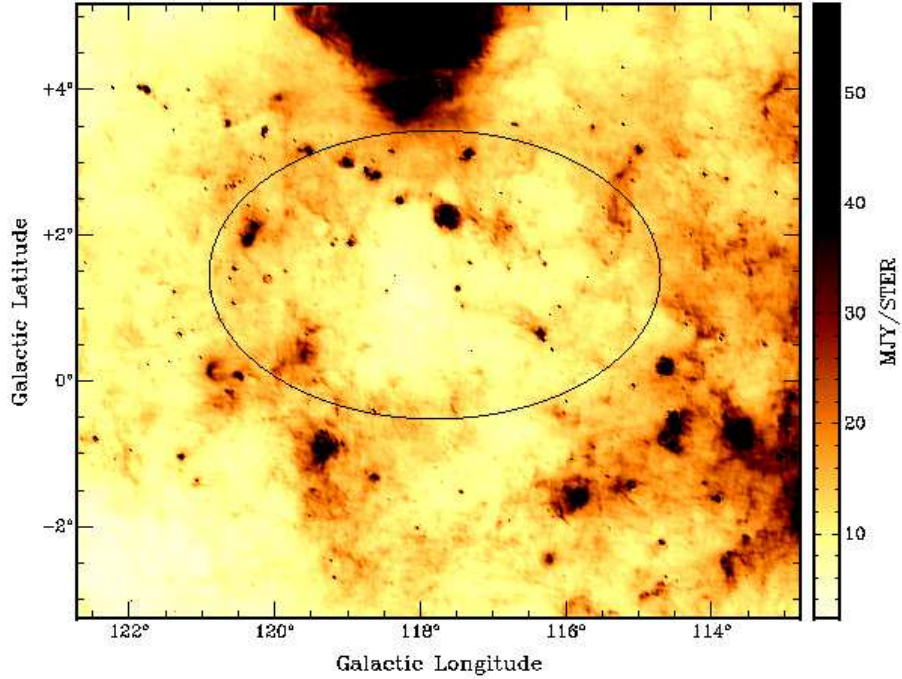}
\caption{60 $\mu$m HIRES IRAS data of the same area as Fig.\ref{gran-shell-1420}.  The ellipse delineates the \hi\ shell shown in Fig.\ref{gran-shell-hi}.}
\label{gran-shell-60}
\end{figure*}

\begin{figure*}
\includegraphics[width=15cm]{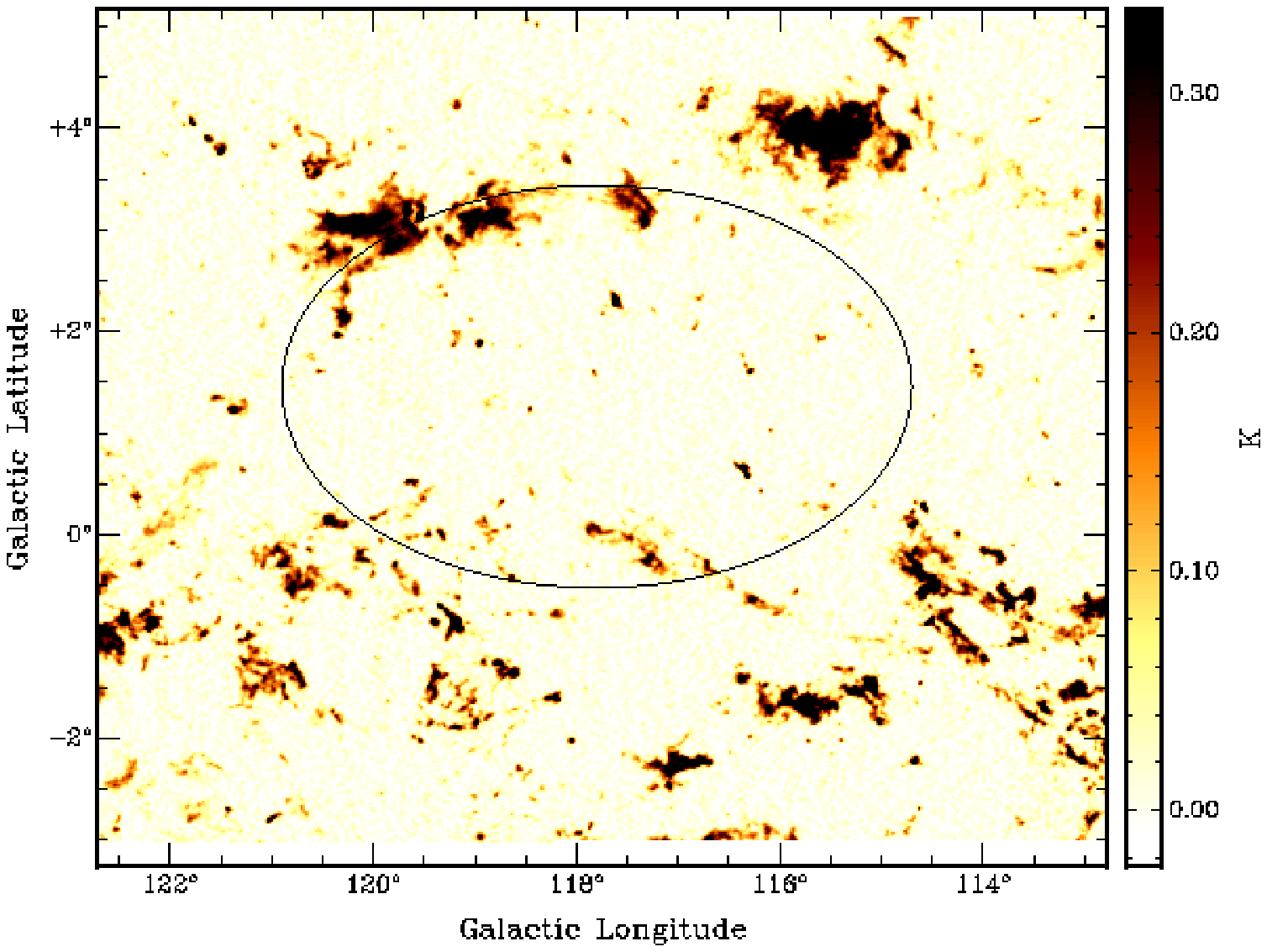}
\caption{ Averaged CO emission distribution in the velocity range from --50 to --20 \kms. The ellipse delineates the \hi\ shell shown in Fig.\ref{gran-shell-hi}.}
\label{gran-shell-co}
\end{figure*}

As mentioned in Section \ref{intro}, Sh2-173 lies in the border of a large \hi\ shell  observed in the Perseus arm \citep{fic86, moo03}. 
Using CGPS data  \citep{tay03}, we have inspected the area looking for the \hi\ structure observed by  \citet{fic86} using low resolution data.

An inspection of the whole data cube reveals the presence of an \hi\ minimum which is first seen near $-20$ \kms\, and which can be followed up to $-50$ \kms. 
Figure \ref{gran-shell-hi} shows  the \hi\ distribution averaged in the velocity range from --20 to --50 \kms, where the 
 large \hi\ cavity is evident centred at ({\it l, b}) = ( 117\fdg8, +1\fdg5), as well as the \hi\ shell encircling it. 
Following the standard nomenclature for galactic shells, we  designate this structure as GSH\,117.8+1.5-35.

 Along the mentioned velocity range, both the \hi\ minimum and the \hi\ shell are well defined as a large structure, although, as in many other \hi\ shells, the typical behavior of an expanding structure  (i.e. a ring that start as a cap, reach its greatest angular dimension at the systemic velocity and then shrink back to a cap) is not completely  observed. In this case the receding cap is missing, while the approaching cap is detected, at about --50 \kms. The absence of the receding cap could be due to confusion effects. Other possible explanation is that the receding part may be expanding into a lower density medium, as it would happen if for example  it were located near the border of the Perseus arm. 
On the other hand, as expected when dealing with high resolution \hi\ data, there are several  small-scale structures inside and around the large ring. These smaller \hi\ features 
change in shape and size along the velocity range where the large shell is defined, and are possibly related to  the individual action of every massive star present in the area. In this paper we will just focus on the large scale structure.

The location of the stars belonging to Cas\,OB5 \citep{gar92} are indicated in Fig. \ref{gran-shell-hi}, showing that most of them lie outside the large 
\hi\ minimum. Particularly, two O-stars belonging to Cas\,OB5 are seen projected towards a smaller \hi\ minimum centred at 
({\it l, b}) = (117\fdg3, -1\fdg3).
The position of the  WN4 star (WR\,159) \citep{neg03} is indicated with a larger circle.

In Figure \ref{gran-shell-1420} a 1420 MHz radio continuum image of the same area displayed in Fig. \ref{gran-shell-hi} is shown. 
The ellipse delineates the location of the \hi\ shell. A minimum in the radio continuum emission in coincidence with the \hi\ cavity 
is observed, although the presence of two SNR (CTB\,1 and G116.5+1.1) makes it less evident.
Several Sharpless \hii\ regions are projected onto the thick \hi\ shell, being Sh2-173 one of them.
Based on H$\alpha$ observations, \citet{fic90} derived the radial velocity of galactic \hii\ regions. From their work, 
we found out that most of the \hii\ regions shown in Fig. \ref{gran-shell-1420} have a radial velocity between --32 and --55 \kms, in 
coincidence with the radial velocity in which the large \hi\ shell is detected. Only two regions, Sh2-171 and Sh2-167, have  radial velocities 
outside this range, of about $-12$ \kms\, and $-66$ \kms\,, respectively. 
As mentioned by \citet{caz03}, Sh2-180, Sh2-181, and Sh2-182, may be associated with Cas\,OB7.
Sh2-179 is a planetary nebula \citep{ker03}.
Besides the \hii\ regions, four SNR are seen projected in the area, G114.3+0.3, G116.5+1.1, CTB\,1, and G120.2+01.4 (Tycho).

The HIRES 60 $\mu$m image of the same area is shown in Fig. \ref{gran-shell-60}. 
 The large-scale structure is similar to the infrared loop GIRL G117+00 shown by \citet{kis04}.
The infrared emission shows a distribution similar to the radio continuum one. A minimum in the 60 $\mu$m emission is evident inside the 
large \hi\, shell (indicated by the ellipse).
One of the main differences between this image and the one shown in Fig. \ref{gran-shell-1420} is the absence of  SNRs at 
infrared wavelengths. As expected, the infrared emission associated with \hii\ regions is clearly observed.

 Figure \ref{gran-shell-co} shows the averaged CO  emission distribution  in the velocity range from --50 to --20 \kms. The CO data were 
obtained from the Five College Radio Astronomical Observatory (FCRAO) CO Survey. The image  reveals several molecular clouds sparsely  located 
at the shell's  border.

It is generally believed that expanding supershells may trigger star formation at their edges. 
 As the shell expands, 
it becomes gravitationally unstable, providing a physical mechanism to form clouds and induce star formation  along its periphery \citep[e.g.][]{mcc87, elm98}.
There are some observational evidence 
confirming the importance of the shell's evolution in creating new stars.
\citet{pat98} analyzed the origin and evolution of the Cepheus bubble and concluded that the Cep\,OB2 association was probably triggered 
by the action of an earlier generation of massive stars and suggested that a third generation of stars has recently formed as a consequence 
of the evolution of the shell. 
\citet{oey05} analyzed the W3/W4 Galactic star forming complex and suggested that it consists of a hierarchical system of three generations: 
a supergiant shell triggered the formation of IC\,1795 in W3, which in turn triggered the star formation in its surroundings. 
They confirmed this scenario based on the age sequence showed for the different generations.
\citet{arn07} found a new \hi\ supershell (GS263-02+45) which has  the OB-association Bochum\,7 located in its border. 
They concluded that the association may have been born as a consequence of the evolution of GS263-02+45.

In our case, having in mind the scenario previously described, i.e. a large expanding \hi\ shell containing several \hii\ regions in its edge, and where at least one of them, Sh2-173, shows evidence of  star forming processes taking place in its surrounding molecular material (see Section \ref{star-formation}), we wonder if this could be other case of a hierarchical system of three generations.
If this were the case, it should be reflected by a corresponding age sequence.
In what follows we will briefly discuss possible age estimates for both, the \hii\ region Sh2-173 and the large \hi\ shell \gsh.

Given that the  main exciting star of Sh2-173 is still visible, an upper limit for its age can be estimated. The main-sequence lifetime for 
an O9 star is about 5.0 Myr \citep{sch97}. On the other hand, a lower limit to the age can be estimated considering the sound crossing time, $R/v_s$, where $v_s$ is the sound speed in the ionized gas ($\sim 15$ \kms) and $R$ is the radius of the \hii\ region. 
For Sh2-173 the corresponding lower limit is about  0.6 Myr. 
 As a different approach, the age of the \hii\ region can be inferred using the simple model described by  \citet{dys80} for an expanding 
ionized region in a uniform medium. 
The expansion of an \hii\ region is highly dependent on the density of the surrounding gas. As a rough estimate,
the original ambient density can be obtained by distributing the total
mass related to the structure (ionized, neutral atomic, and molecular)
over a sphere of  12 $\pm$ 2 pc in radius, which yields $n_o =  41 \pm 25 $ cm$^{-3}$.
Under  these conditions, we can infer that the \hii\ region has been expanding during 0.1 - 1.0 Myr. This range for the dynamical age, 
together with the age limits estimated above, suggest that 
 the age of Sh2-173 could be between  0.6 and 1.0 Myr.
 Using the same simple model the expansion velocity of the \hii\ region can be inferred. 
We obtained $V_{\rm exp} = 6 \pm 3$ \kms, which is in agreement with the expansion velocity estimated for the molecular gas (4 $\pm$ 2 \kms). 
Thus, the ionized and molecular gas are expanding at the same velocity, as  expected according to the models of \citet{hos06} when the molecular material has been accumulated behind the ionization front.

A  rough estimate for the  age of the large \hi\ shell, \gsh\,,  can be obtained using a simple model to describe the expansion of a 
shell created by a continuous injection of mechanical energy \citep{wea77}. In this way,  the dynamical age of the \hi\ structure can be 
estimated as $t_{\rm dyn} = 0.6\, R$ (pc)$ /V_{\rm exp}$(\kms), where $R$ is the radius of the shell and $V_{\rm exp}$ the expansion velocity. 
 The expansion velocity can be obtained as $V_{\rm exp} = (v_2 - v_1) /2  + 1.3$ \kms\, (see Section \ref{hi}). Given that the large 
\hi\ shell is observed 
over some 30 \kms, an expansion velocity of about 16 \kms\, is inferred.
Thus, considering $R = 135 \pm 20$ pc (for a distance of 2 kpc, see the discussion below) and $V_{\rm exp} = 16 \pm 2$ \kms, an age 
of 5 $\pm$ 1 Myr is obtained, which is large as compared with the age estimate obtained for Sh2-173.

Concerning the origin of the large shell, \citet{fic86} suggested that the three supernova remnants located within the \hi\ shell have 
actively  enlarged it, which was presumably previously created by the joined action of several stellar winds and/or several supernova remnants. 
Based on the location of Cas\,OB5 on the sky, \citet{moo03}  associated the expanding shell with the OB association. However, an energetic 
analysis has not been made in order to test this possibility. Are the massive stars (O-type stars) belonging to Cas\,OB5 capable of creating 
such a shell?.
The first condition that should be checked is if the \hi\ shell and Cas\,OB5 are at comparable distances from  the Sun. 
The systemic velocity adopted for the large shell is $-35$ \kms\,, while according to \citet{hum76}, Cas\,OB5 has $V_{\rm LSR} = -39.5$ \kms.
On the other hand, \citet{gar92} determined for Cas\,OB5 a distance modulus of DM = 11.5 mag, which yields a distance  D = 2 kpc.
As mentioned in Section \ref{intro},  the presence of non-circular motions in the Perseus arm is important and should be taken into account.
An inspection of the {\it observed} radial velocity field of the Galaxy derived 
by \citet{bra93} gives for the adopted systemic velocity a distance of about 1.8-2 kpc.
Thus, we will consider a distance of 2 kpc for Cas\,OB5 and the \hi\ shell.

A second condition that should be checked is if the energy budget of the massive stars is enough to create the large shell. 
The mechanical energy output of the stars should be compared  with the kinetical energy of the shell.
As can be seen in Fig. \ref{gran-shell-hi}, 
 four O-type stars (BD+62\,2299 (O8), BD+61\,2550 (O9.5II), BD+61\,2559 (O9V), and HD\,108 (O5 f))
 and one Wolf-Rayet star (WR\,159 (WN4)) are seen projected inside the large \hi\ minimum.
All these massive stars are supposed to belong to Cas\,OB5 \citep{gar92,neg03}.
The total mechanical energy released by the stars during a time $t$ is $E_w = (\Sigma 0.5\, \dot{M}\ v^2_{\infty})\,t$, 
where $\dot{M}$ is the mass-loss rate and $v_{\infty}$ the terminal velocity of the winds.  The corresponding values were 
taken from \citet{how89} for the O-type stars
 (log\, $\dot{M}$ = --5.73, --6.76, --7.17, and --7.36; and  
$v_{\infty}$ = 2800, 2200, 2100, 
and 2000 \kms\, for the O5, O8, O9, and O9.5 stars, respectively)  and from \citet*{van86} and \citet*{pri90} for the WR star
  (log\, $\dot{M}$ = --4.5 and $v_{\infty}$ = 1900 \kms).
We obtained that the total energy input due to the stellar winds is about $E_w = 1.3 \times 10^{45}\,t(\rm yr)$ erg. 

The total kinetic energy stored in the shell is $E_k = 0.5\, M_{\rm sh}\,V_{\rm exp}^2$, where $M_{\rm sh}$ is the total mass in the shell 
and $V_{\rm exp}$ the expansion velocity. We estimated the expansion velocity of the shell to be 16 $\pm$ 2 \kms. Following the procedures 
described in Sections \ref{hi} and \ref{co}, we obtained the \hi\ and molecular masses in the shell.  Each mass was computed from  the corresponding  averaged image covering the velocity range from --20 to --50 \kms.
Adopting for \gsh\, a distance of 2  kpc, the total swept up mass is $M_{\rm sh} = (8 \pm 4) \times 10^5 M_{\odot}$,  which is consistent 
with the value estimated by \citet{moo03} of 7.5 $\times 10^5 M_{\odot}$  (for their adopted distance of 2.5 kpc, which implies a 
mass of 4.8 $\times 10^5 M_{\odot}$ for a distance of 2 kpc).
Thus, the kinetic energy of the shell is $E_k = (2 \pm 1) \times 10^{51}$ erg. 

The theoretical model of \citet{wea77} estimates that the efficiency of conversion of mechanical wind energy $E_w $ into kinetic energy of the 
shell $E_k$ is up to 20 \% for the energy conserving model. This implies that the massive stars have to be acting during at least 8 Myr to
 create the shell. This value is larger than the lifetime of an O-type star and suggests that the five massive stars  can not alone generate 
such a shell and other sources of energy should be considered. As suggested by \citet{fic86}, the supernova remnants located within the \hi\ 
shell may be actively enlarging it.
Out of the four SNRs located in the area, G114.3+0.3 is projected outside the \hi\ shell and it is located at a distance of about 700 pc 
\citep*{yar04}. G114.3+0.3 and CTB1 have a distance estimate of about 1.6 kpc \citep{yar04}, while Tycho is located at about 2.4 kpc 
\citep*{lee04}.
Thus, assuming that there are three SNR possibly related to the shell and that each of them  has injected $10^{51}$ erg, then the massive 
stars have to be acting during at least 5 Myr, which is consistent with the dynamical age estimated for the large shell.
It is important to note that given the large errors involved in these estimations, the possibility that other sources of energy (stellar winds
 and/or supernova explosions) would have contributed in the formation of the shell can not be discarted. For instance, it is highly probable 
that the progenitors of the three SNR have contributed in the creation of the shell.

All the evidence shown above  suggests that Sh2-173 is part of a hierarchical system of three generations. The members of Cas\,OB5 would have formed the large \hi\ shell, whose expansion would have triggered the formation of the \hii\ region Sh2-173, which in turn is triggering new stars in the surrounding molecular material.  
The fact that  Sh2-173 is one of several  young \hii\ regions lying at the edge of the \hi\ shell (see Fig.\ref{gran-shell-1420}) reinforces 
this conclusion.  For example, Sh2-170  has been analyzed by \citet{rog04}, who derived an age of $(2.5 - 5) \times 10^5$ yr, while Sh2-168 is 
considered as a compact \hii\ region by \citet{kot02}. 
A detailed study of the  \hii\ regions shown in Fig. \ref{gran-shell-1420} would be useful to better understand the importance that the 
expansion of the large \hi\ shell had in the process of triggering the formation of new generations of massive stars.

\section{Conclusions}

The \hii\ region Sh2-173 has been analyzed in order to study the phenomena associated with the interaction of massive stars and the interstellar medium, and specifically, the process of triggering star formation.
The analysis of the available radio continuum data have enabled us to estimate some parameters that characterize the \hii\ region, like the ionized mass, the emission measure, and the electron density. From the radio flux densities estimated at different wavelengths, the thermal nature of the source was confirmed by the estimation of the spectral index $\alpha = 0.0 \pm 0.1, S_{\nu} \sim \nu^{\alpha}$. We noted that the massive stars present in the field can provide the necessary UV photon flux to keep the \hii\ region ionized and heat the dust.
As expected for \hii\, regions, Sh2-173 displays a strong correlation between the 1420 MHz brightness temperature and infrared emission.
An inspection of the \hi\ images has shown the existence of an \hi\ cavity which presents a good morphological agreement with the ionized gas. The \hi\ feature is observed within the velocity range --24.6 and --31.6 \kms\, and has a systemic velocity of $-27.0 \pm 1.3$ \kms\,, compatible with the velocity of the ionized gas as obtained from H$\alpha$ observations. 
The distribution of the  MSX emission at 8.3 $\mu$m clearly shows the presence of a PDR bordering the brightest part of the \hii\ region.
Molecular gas probably related to Sh2-173 is observed in the velocity range  --27.4 to --40.6 \kms. Particularly, the CO feature named cloud A is clearly observed encircling both the \hii\ region and the PDR.
Applying  different colour criteria, 46 YSO candidates have been identified  in the vicinity of Sh2-173. Their positions in the CC and CM diagrams, as well as their correlation with molecular emission highly suggest the presence of an active star formation process in the molecular envelope of Sh2-173.

An analysis of the \hi\ distribution in a larger area shows that Sh2-173 is located at the edge of a large shell, together with several other \hii\ regions. This morphology suggests that Sh2-173 may be part of a hierarchical system of three generations. In this context, the age of the different structures were estimated.
Based on evolutionary models of \hii\ regions, we estimated for Sh2-173 a dynamical age of  0.6 - 1.0 Myr.
As for the large \hi\ shell, we have obtained a dynamical age $t_{\rm dyn} = 5 \pm 1 $ Myr.
Based on energetics considerations, we have concluded that the massive stars belonging to Cas\,OB5 could not alone create the shell, being highly probable that at least three supernova explosions have contributed in its formation. 
The obtained age difference between Sh2-173 and the \hi\ shell, together with  their relative location, lead us to the conclusion that 
Sh2-173 may have been created as a consequence of 
  the  action of a strong shock produced by the expansion of \gsh\, into the surrounding molecular gas. The analysis
of the ages and distribution of the other \hii\ regions observed in the area would be useful to confirm this scenario.

\section*{acknowledgments}{
We would like to acknowledge Dr. E.M. Arnal for very useful comments and suggestions.
This research has made use of the NASA/ IPAC Infrared Science Archive, which is operated by the Jet Propulsion Laboratory, California Institute of Technology, under contract with the National Aeronautics and Space Administration.
The CGPS is a Canadian project with
international partners and is supported by grants from NSERC.
Data from the CGPS
is publicly available through the facilities of the Canadian
Astronomy Data Centre (http://cadc.hia.nrc.ca) operated by the 
Herzberg Institute of Astrophysics, NRC. The WENSS project is a
  collaboration between the Netherlands Foundation for Research in
  Astronomy (NFRA/ASTRON) and the Leiden Observatory WENSS team. 
This research has made use of the SIMBAD database,
operated at CDS, Strasbourg, France.
This project was partially financed by the Consejo Nacional de Investigaciones Cient\'{\i}ficas y T\'ecnicas (CONICET) of Argentina under projects PIP 5886 and PIP 6433, Agencia PICT 00812, UBACyT X482, and UNLP 11/G072.
We are grateful to the referee, whose suggestions led to the improvement of this paper. }

\bibliographystyle{mn2e}  
\bibliography{sh173-biblio}

\IfFileExists{\jobname.bbl}{}
{\typeout{}
\typeout{****************************************************}
\typeout{****************************************************}
\typeout{** Please run "bibtex \jobname" to optain}
\typeout{** the bibliography and then re-run LaTeX}
\typeout{** twice to fix the references!}
\typeout{****************************************************}
\typeout{****************************************************}
\typeout{}
}

\end{document}